\documentclass[final,3p,times,twocolumn,longtitle]{elsarticle}

\usepackage{color}
\usepackage{multirow}
\usepackage{braket}
\usepackage{bm}
\usepackage{hyperref}
\usepackage{tabularx}
\usepackage{lineno}
\usepackage{amssymb}
\usepackage{amsmath}
\usepackage{subfigure}
\usepackage{amsthm}

\newcommand{\Edep} {E$_{\mathrm {dep}}$}
\newcommand{\Evis} {E$_{\mathrm {vis}}$}
\newcommand{\Erec} {E$_{\mathrm {rec}}$}

\newcommand{\B}{$^{12}$B}
\newcommand{\N}{$^{12}$N}
\newcommand{\Tl}{$^{208}$Tl}
\newcommand{\Bi}{$^{212}$Bi}
\newcommand{\BI}{$^{214}$Bi}

\newcommand{\tabincell}[2]{\begin{tabular}{@{}#1@{}}#2\end{tabular}}
\newcommand{\electron}{$e^-$}
\newcommand{\positron}{$e^+$}
\newcommand{\elecposi}{$e^\pm$}

\definecolor{darkgreen}{rgb}{0.0, 0.5, 0.0}
\definecolor{orange}{rgb}{1, 0.5, 0.0}

\newcommand{\nuebar}{$\overline{\nu}_{e}$}
\journal{Nucl. Instr. Meth. A}

\graphicspath{{figures/}}

\begin{document}

\begin{frontmatter}

\title{A high precision calibration of the nonlinear energy response at Daya Bay}

\def\IHEP{1}
\def\ECUST{2}
\def\Wisconsin{3}
\def\Yale{4}
\def\BNL{5}
\def\NTU{6}
\def\NJU{7}
\def\NUU{8}
\def\TsingHua{9}
\def\SZU{10}
\def\ZSU{11}
\def\NCEPU{12}
\def\CUHK{13}
\def\Dubna{14}
\def\Siena{15}
\def\USTC{16}
\def\Charles{17}
\def\UIUC{18}
\def\LBNL{19}
\def\BNU{20}
\def\XJTU{21}
\fntext[BCC]{Now at: Department of Chemistry and Chemical Technology, Bronx Community College, Bronx, New York  10453}
\def\UH{23}
\def\CIAE{24}
\def\SDU{25}
\def\VirginiaTech{26}
\def\NCTU{27}
\def\UC{28}
\def\TempleUniversity{29}
\def\DGUT{30}
\def\UCB{31}
\fntext[UPENN]{Now at: University of Pennsylvania, Department of Physics \& Astronomy, 209 South 33rd Street, Philadelphia, PA 19104-6396}
\def\HKU{33}
\def\NanKai{34}
\def\IIT{35}
\def\SJTU{36}
\def\Princeton{37}
\def\CalTech{38}
\def\WM{39}
\def\CUC{40}
\def\UCI{41}
\def\CGNPG{42}
\def\NUDT{43}
\def\IowaState{44}
\def\CQU{45}
\author[\IHEP]{D.~Adey}
\author[\ECUST]{F.~P.~An}
\author[\Wisconsin]{A.~B.~Balantekin}
\author[\Yale]{H.~R.~Band}
\author[\BNL]{M.~Bishai}
\author[\NTU]{S.~Blyth}
\author[\NJU]{D.~Cao}
\author[\IHEP]{G.~F.~Cao}
\author[\IHEP]{J.~Cao}
\author[\IHEP]{J.~F.~Chang}
% ...

\begin{abstract}
\noindent A high precision calibration of the nonlinearity in the energy response of the
Daya Bay Reactor Neutrino Experiment's antineutrino detectors is presented in detail.
The energy nonlinearity originates from the particle-dependent light yield of the
scintillator and charge-dependent electronics response.
The nonlinearity model is constrained by $\gamma$~calibration points from deployed and naturally occurring radioactive sources, the $\beta$~spectrum from \B~decays,
and a direct measurement of the electronics nonlinearity with a new flash analog-to-digital converter readout system. Less than 0.5\% uncertainty in the energy nonlinearity calibration is achieved for positrons of kinetic energies greater than 1~MeV.
\end{abstract}


\end{frontmatter}


\section{Introduction}
\label{Introduction}

Liquid scintillator~(LS), together with photosensors such as photomultiplier tubes~(PMTs),
has been widely used for energy calorimetry.
This kind of detector plays an important role in neutrino physics,
especially in reactor \nuebar~experiments, from the discovery of the neutrino at the Savannah River Plant~\cite{Cowan}, to the measurement of the neutrino oscillation angle $\theta_{12}$~at KamLAND~\cite{KamLAND2008}, the observation of the neutrino oscillation driven by $\theta_{13}$~at Daya Bay~\cite{DYBPRL2012}, RENO~\cite{RENO2012}, and Double Chooz~\cite{DC2012}, and the next generation experiment such as JUNO~\cite{JUNOYB}. 

With reactor antineutrinos in the MeV energy range, these experiments are sensitive to the \nuebar~disappearance. The survival probability is expressed as
\begin{equation}
 \label{eq:oscillation}
\begin{split}
&P_{\rm survival} = 1 - \cos^4\theta_{13}\sin^2 2\theta_{12} \sin^2\Delta_{21} \\
  &- \sin^2 2\theta_{13}(\cos^2 \theta_{12} \sin^2 \Delta_{31} + \sin^2 \theta_{12} \sin^2\Delta_{32}),
\end{split}
\end{equation}
where $\Delta_{ij} \simeq 1.267 \Delta m^{2}_{ij} {\rm L}/{\rm E}_\nu$, L in meters is the reactor-detector distance, E$_\nu$~in MeV is the \nuebar~energy,
and $\Delta m^{2}_{ij}$~in ${\rm eV^2}$~is the mass-squared difference between neutrinos of mass eigenstates $\nu_i$ and $\nu_j$. Reactor antineutrinos are detected via the inverse beta decay reaction~(IBD) \nuebar$+p\to e^++n$.
The \positron~and neutron form a prompt and delayed coincident pair which efficiently discriminates \nuebar~interactions from background. Convolving the \nuebar~energy spectrum with the IBD cross section~\cite{IBDCrossSection} results in an expected spectrum which rises from the 1.8~MeV interaction threshold, peaks at $\sim$4~MeV, and falls to a very low rate above 8~MeV.

Based on the kinematics of IBD reactions, the neutron generally carries only a small fraction of the initial \nuebar~energy, O(10~keV). The initial \nuebar~energy can be estimated using ${\rm E}_\nu \simeq {\rm E}_{\rm prompt} + 0.78$~MeV, where E$_{\rm prompt}$~is the sum of the positron kinetic plus annihilation energy. 
Since the observable in the detector is the reconstructed positron energy E$^{e^+}_{\mathrm {rec}}$, the precise understanding of the detector energy response, which connects E$_{\rm prompt}$~and E$^{e^+}_{\mathrm {rec}}$, is critical for the measurement of the oscillation parameters and for the absolute measurement of the reactor \nuebar~spectrum.

The energy response of LS is not linear with respect to the kinetic energies of charged particles, primarily due to ionization quenching~\cite{BirksBook} and Cherenkov radiation~\cite{Cherenkov}.
The former reduces the scintillation light yield at high ionization density, such as protons, $\alpha$~particles and low energy \elecposi,
while the latter contributes additional photons if a particle's velocity is larger than the phase velocity of light in the LS. Further nonlinearity can be introduced during the PMT charge estimation in the electronics readout system. 

In this paper, we present a precise energy nonlinearity calibration of the Daya Bay antineutrino detectors~(ADs). An energy nonlinearity model is built accounting for the scintillation and electronics nonlinearity. The model is constrained by ten $\gamma$~calibration points, the $\beta$~spectrum from \B~decays,
and a direct measurement of the electronics nonlinearity with a newly installed flash analog-to-digital converter~(FADC) readout system. Less than 0.5\% uncertainty in the energy nonlinearity calibration is achieved for positrons of kinetic energies greater than 1~MeV.

\section{The Daya Bay experiment}
\label{DayaBay}
\begin{figure}[!htb]
\begin{centering}
\includegraphics[width=.45\textwidth]{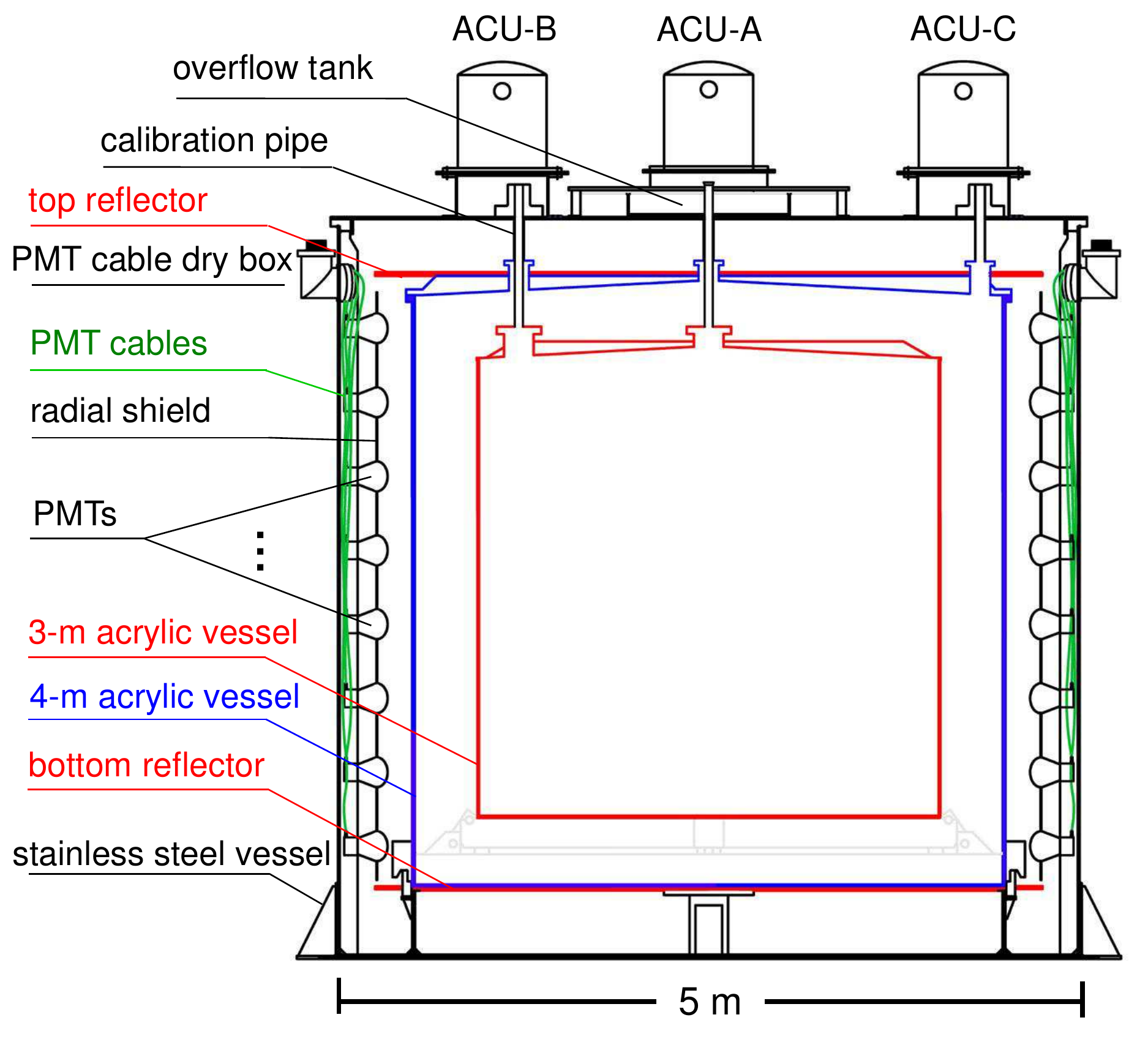}
\caption{\label{fig:AD} Schematic of a Daya Bay antineutrino detector.
The three cylindrical volumes, defined by three acrylic vessels, are filled with Gd-LS, LS, and mineral oil
from the innermost to the outermost.
Three automatic calibration units are installed on top of the detector to calibrate the PMTs and detector energy response.}
\end{centering}
\end{figure}

The Daya Bay experiment consists of eight identically designed antineutrino detectors~(ADs). Two of them are deployed in each of two near halls~(EH1 and EH2) and four are installed in the far hall~(EH3). To designate a particular AD, EH1-AD1 means AD1 in EH1, for example. 
Each AD has three nested cylindrical volumes separated by concentric acrylic vessels as shown in Fig.~\ref{fig:AD}.
The innermost volume is filled with 20~tons of gadolinium-loaded LS~(Gd-LS), serving as the primary \nuebar~target.
It is surrounded by $\sim$22~tons of non-loaded LS to detect $\gamma$~rays escaping from the target volume.
The outermost volume is filled with mineral oil to shield the LS from natural radioactivity.
A total of 192 8-inch PMTs are installed on the steel vessel~(SS) to detect scintillation photons,
giving about 160~photoelectrons per MeV of deposited energy~(p.e./MeV).
The ADs are immersed in a water Cherenkov muon detector which functions both as a passive radiation shield and as an active muon identification. Details of the detector system are described in Refs.~\cite{DYBDetector,DYBMuon}.

There are three automatic calibration units~(ACUs) on the top of each AD to calibrate the energy response
along the vertical axes at detector center~(ACU-A), the edge of the Gd-LS volume~(ACU-B),
and the LS volume~(ACU-C).
Regular calibration consists of the deployment of a $^{68}$Ge source, a $^{60}$Co source,
a $^{241}$Am-$^{13}$C neutron source~\cite{DYBAMC}, or a light-emitting diode~(LED).
In the summer of 2012, a special calibration deployed various radioactive sources to calibrate the
energy nonlinearity~($^{137}$Cs, $^{54}$Mn, $^{40}$K, Pu-$^{13}$C, $^{241}$Am-$^{9}$Be). 
Details of the calibration system are described in Ref.~\cite{DYBACU}.

\begin{figure}[!htb]
\begin{centering}
\includegraphics[width=.45\textwidth]{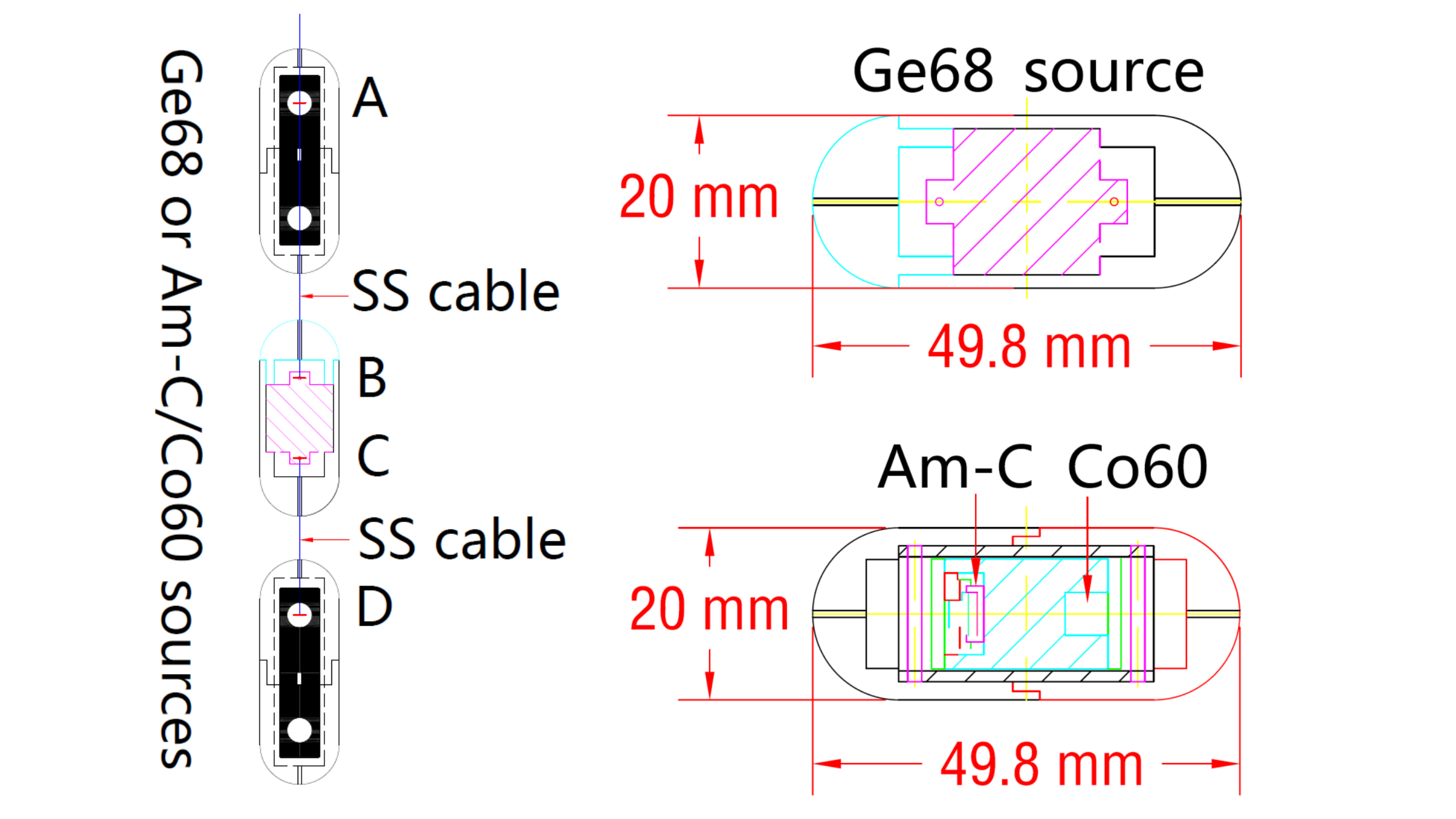}
\caption{\label{fig:source} Left: schematic for the radioactive sources,
including a top and a bottom weight~(A and D) to maintain tension in the deployment string~\cite{DYBACU}.
Right: sections of the $^{68}$Ge, $^{241}$Am-$^{13}$C and $^{60}$Co sources. Dimensions are millimeters.}
\end{centering}
\end{figure}

The radioactive sources are housed in a stainless steel capsule to shield the accompanying \electron~or \positron~from the scintillator.
The capsule is further enclosed by acrylic, polytetrafluoroethylene~(PTFE) with high reflectivity or greenish PTFE as shown in Fig.~\ref{fig:source}. In the following the greenish PTFE will be referred to as greenish teflon. There are weights on the top and bottom of the source, made of stainless steel surrounded by acrylic,
to maintain tension in the deployment string. In January 2017, another special calibration campaign was performed to better understand the optical shadowing of the source enclosures and deployment system, which will be discussed in Sec.~\ref{GammaData}.

The scintillation photons are detected by 192 PMTs~(Hamamatsu R5912). The PMT signals share
the same cable with the high voltage power. A passive input to an AC circuit
is used to decouple the fast PMT signals, which are sent to
the frontend electronics~(FEE)~\cite{DYBFEE}. After an initial fast amplification, the signal is fed
to a pulse shaping circuit consisting of a differential CR and four integrating RC circuits, and then amplified by a factor of ten. The peak value of the shaped PMT signal is used as an estimate of the PMT charge output, and it is sampled by a 40-MHz 12-bit ADC.
 
The energy reconstruction in data starts with a scale conversion from PMT charge~(${\rm Q}_{i}$) to number of photoelectrons,
with a daily calibrated conversion factor of roughly 19~ADC counts per single p.e.~($\overline{{\rm Q}}^{\rm SPE}_{i}(t)$). The estimate of the total energy deposited by a particle interaction, \Erec, is proportional to the total number of photoelectrons over the active PMTs~($\sum_i \frac{{\rm Q}_{i}}{\overline{{\rm Q}}^{\rm SPE}_{i}(t)}$), as
\begin{equation} \label{eq:recE}
  {\rm E}_{\rm rec} = \left(\sum_i \frac{{\rm Q}_{i}}{\overline{{\rm Q}}^{\rm SPE}_{i}(t)}\right) \cdot \frac{f_{\rm act}(t)}{N_{\rm PE}(t)} \cdot f_{\rm pos}(\mathbf{r}_{\rm rec}, t),
\end{equation}
where the correction $f_{\rm act}(t)$, defined as the ratio of total to active AD PMTs, is to compensate for the reduced light when a PMT channel is temporarily disabled\footnote{The average over all ADs of the number of disabled channels is less than 0.5.}. 
The time dependent light yield per MeV of deposited energy, $N_{\rm PE}(t)$, is a scale conversion calibrated by the
neutron capture on Gd with a mean energy of 8.05~MeV. 
The time dependent nonuniformity correction, $f_{\rm pos}(\mathbf{r}_{\rm rec}, t)$,
compensates for the observed variation in total p.e.~versus the estimated position $\mathbf{r}_{\rm rec}$ in the AD.
The light yield and nonuniformity correction are determined by two independent methods: calibration sources and spallation neutron captures. Details can be found in Ref.~\cite{DYBnGd2016}.

\section{The energy nonlinearity model}
\label{EnergyResponse}

In the \nuebar~analysis of Daya Bay~\cite{DYBnGd2016,DYBSpec2017,DYBOsc2018}, the detector energy response
is analytically described in three steps connecting four quantities: 1) the \positron~true energy ${\rm E}_{\rm prompt}$, which is the sum of the positron kinetic plus annihilation energy, 2) the \positron~deposited energy in the Gd-LS and LS \Edep, 3) the visible energy proportional to the total number of detectable photons \Evis, and 4) the reconstructed energy \Erec. The first step is described by a matrix generated with the Geant4~\cite{Geant4} based simulation and is reported in Ref.~\cite{DYBSpec2017}. 
The nonlinear relationship between \Edep~and \Erec, expressed as $f_{\rm NL}$ = \Erec/\Edep, has two origins: in the second step, the total detectable light is not linear with respect to the total deposited energy, and in the third step, the total reconstructed PMT charge is not linear to the total number of photoelectrons generated by the PMT photocathodes.

To decouple the two origins, the visible energy \Evis~is introduced and the nonlinearity expression is rewritten as $f_{\rm NL} = \frac{{\rm E}_{\rm vis}}{{\rm E}_{\rm dep}}\times \frac{{\rm E}_{\rm rec}}{{\rm E}_{\rm vis}}$. In this way the particle-dependent scintillation nonlinearity is described by $\frac{{\rm E}_{\rm vis}}{{\rm E}_{\rm dep}}$, while the charge-dependent electronics nonlinearity is described by $\frac{{\rm E}_{\rm rec}}{{\rm E}_{\rm vis}}$. We will first discuss the modeling of scintillation nonlinearity in Sec.~\ref{ScintNL} before introducing the direct measurement of the electronics nonlinearity in Sec.~\ref{ElectronicsNL}.

It should be noted that the step-by-step strategy is feasible only in the case of negligible coupling between energy leakage and energy nonlinearity, such as the \nuebar~analysis based on the neutron capture on gadolinium. This approach cannot be used in an experiment with significant energy leakage.

\subsection{Scintillation nonlinearity}
\label{ScintNL}

The key particles involved in the \nuebar~detection with LS are \electron, \positron~and the gammas~($\gamma$s). The modeling of their scintillation nonlinearity is discussed in this section.

\subsubsection{Scintillation nonlinearity model of the \electron}
\label{ElectronNL}

\paragraph{Ionization quenching}

When a particle deposits energy in the Gd-LS or LS,
molecules of the solvent~(linear alkylbenzene in Daya Bay, referred to as LAB)~are excited.
The energy is transferred to fluorescent molecules~(3~g/L PPO) primarily via dipole-dipole interactions~\cite{EnergyTransfer}.
Some fraction of energy is not transferred due to quenching of the excitation by the high density
of ionized and excited molecules~\cite{BirksBook}.
The consequence is that the scintillation light yield is not proportional to the deposited energy
for the highly ionizing particles such as protons, $\alpha$s, and low energy \elecposi.

There are many measurements of the quenching effects in various kinds of the organic
scintillator~\cite{FeihongMeasurement,DCMeasurement,BNLMeasurement,KamLANDMeasurement}.
The Birks' model, expressed as
\begin{equation}
 \label{eq:BirksLaw}
 {\rm E}_{\rm scint}~({\rm E}_{\rm dep},~kB) = \int_{0}^{{\rm E}_{\rm dep}}\frac{dE}{1+kB\times \frac{dE}{dx}} ,
\end{equation}
is a popular way to describe quenching,
where ${\rm E}_{\rm scint}$~is the energy converted to scintillation photons, $kB$~is the Birks' coefficient,
and $dE/dx$~is the stopping power.

Equation~\ref{eq:BirksLaw} can be calculated with either a numerical integral or a Geant4 simulation.
The former requires knowing $dE/dx$ as a function of energy,
which is obtained from an ESTAR calculation~\cite{ESTAR} using the Daya Bay LS properties.
For the latter, the Birks' model is implemented in each simulation step when converting the deposited energy
to scintillation photons.

A key configurable in the simulation is the production threshold of secondary particles.
For example, in the Daya Bay simulation the default threshold of \electron~is 0.1~mm~($\sim$80~keV).
If a secondary \electron,
namely a $\delta$~electron, was generated with a kinetic energy less than $\sim$80~keV,
it would not be tracked by Geant4, so the energy loss due to the $\delta$ electron~is added to the ionization loss.

\begin{figure}[!htb]
\begin{centering}
\includegraphics[width=.45\textwidth]{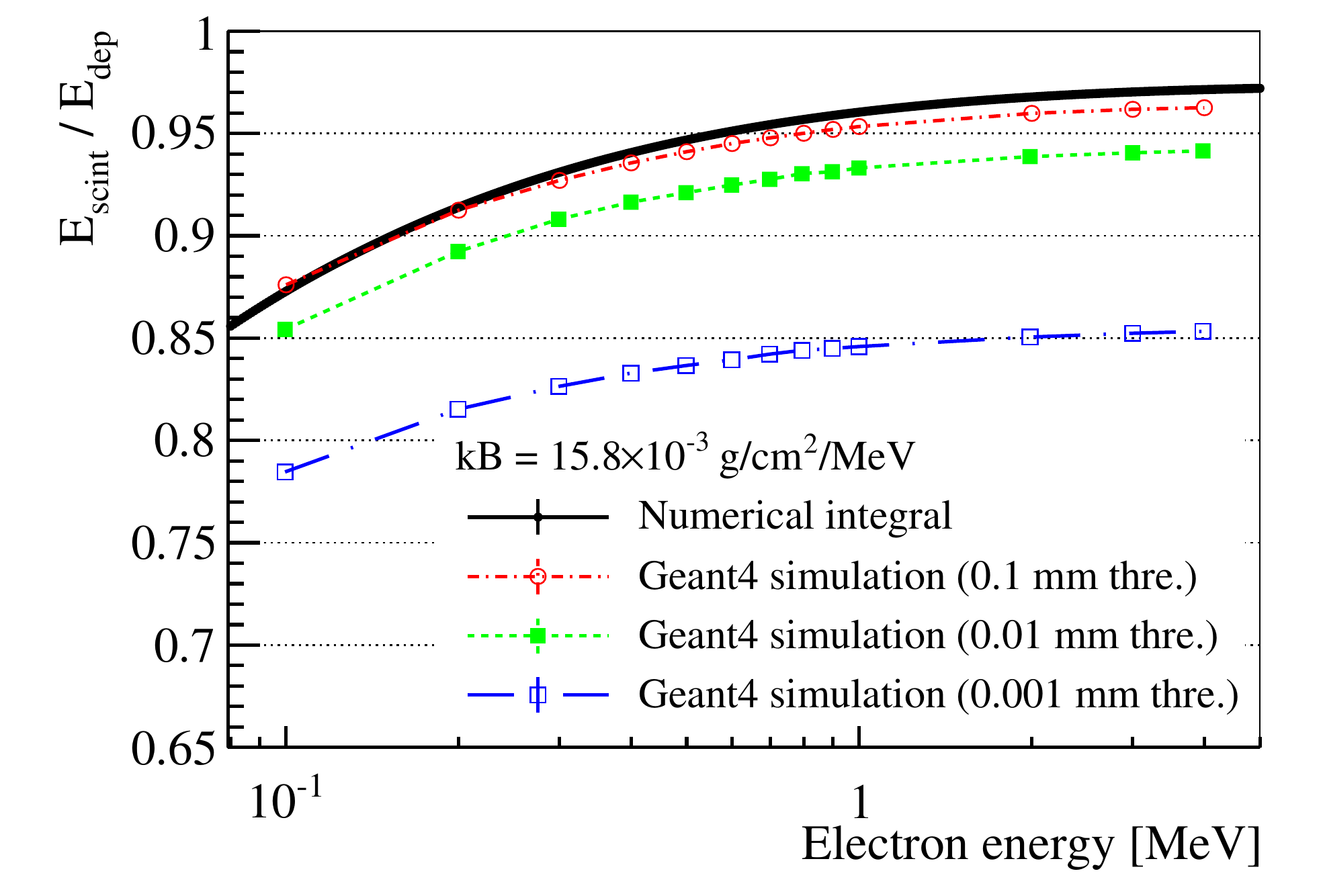}
\caption {\label{fig:ScintillatorQuenching} Relationship between the deposited energy \Edep~and the energy converted to scintillation photons ${\rm E}_{\rm scint}$, derived from the numerical calculation with the ESTAR $dE/dx$~(black),
the Geant4-based Daya Bay simulation with a 0.1~mm particle production threshold~(red), the simulations with 0.01~mm~(green) and 0.001~mm~(blue) production threshold. The lower threshold yields more $\delta$~electrons, resulting in a heavier quenching. }
\end{centering}
\end{figure}

\begin{figure}[!htb]
\begin{centering}
\includegraphics[width=.45\textwidth]{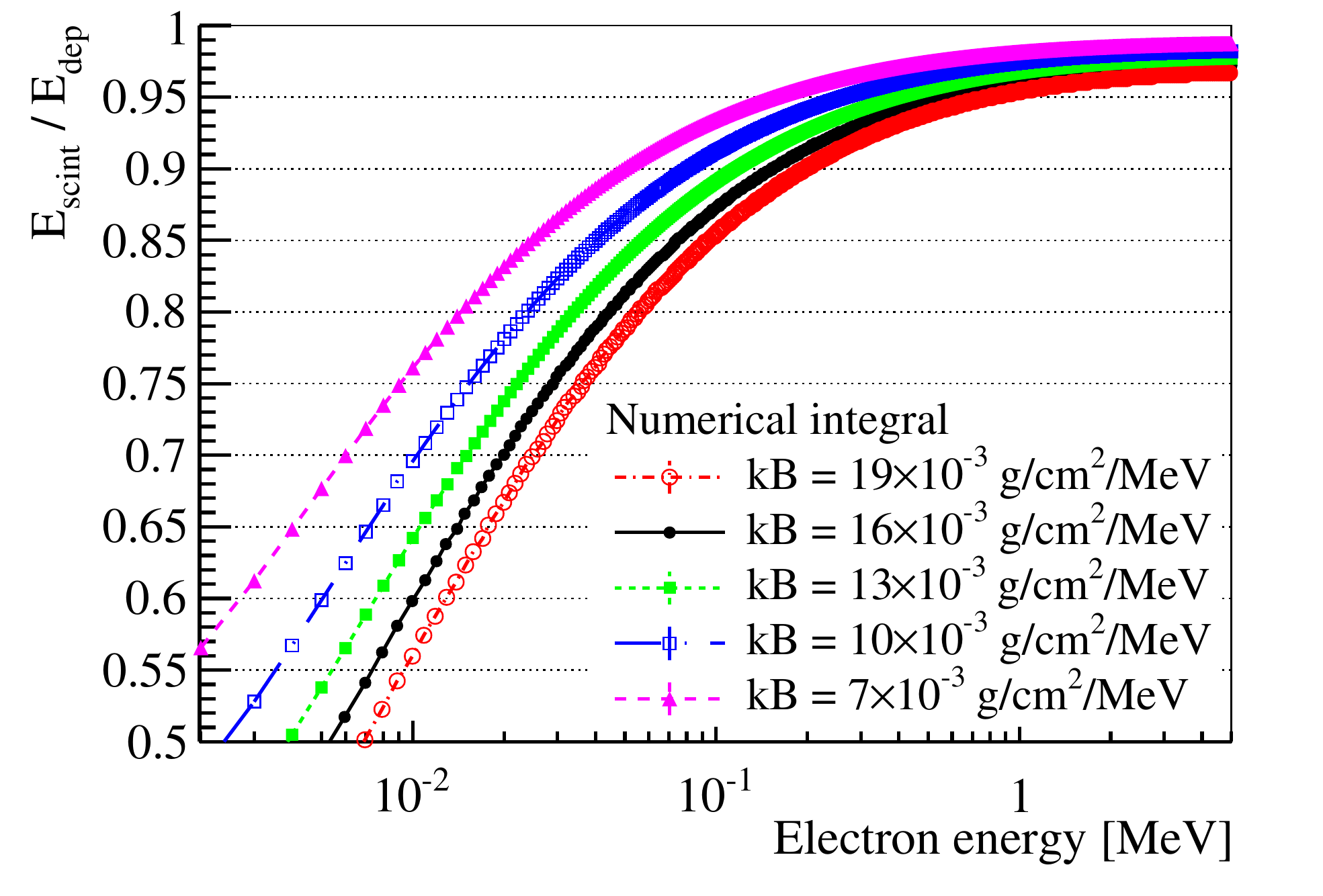}
\caption {\label{fig:ScintillatorQuenching_1} Relationship between \Edep~and ${\rm E}_{\rm scint}$~with different $kB$ values, calculated with the numerical integral method. The larger $kB$~yields the larger quenching.}
\end{centering}
\end{figure}

Thus there are several strategies to generate the quenching curves, which describe the relationship between \Edep~and ${\rm E}_{\rm scint}$.
Figure~\ref{fig:ScintillatorQuenching} shows four curves generated with different strategies:
a numerical integral with $dE/dx$ from ESTAR, a simulation with the default production threshold 0.1~mm,
and two other simulations with 10 and 100 times lower production thresholds, i.e., 0.01~mm and 0.001~mm, respectively.
Even with the same $kB$, the quenching factors from those methods do not agree.
The lower the production threshold, the more $\delta$~electrons are produced, resulting in heavier quenching.
This means that the $kB$ values reported by different measurements cannot be compared without a unified simulation configuration, as similarly reported in Ref.~\cite{DCMeasurement}.
Figure~\ref{fig:ScintillatorQuenching_1} shows the quenching curves with different $kB$~values. A larger $kB$~yields the larger quenching.
It is difficult to determine which calculation method is better, thus
the numerical integration method is adopted as the default and cross validations were performed by simulated quenching curves with different thresholds.

\paragraph{Cherenkov radiation}

\begin{figure}[!htb]
\begin{centering}
\includegraphics[width=.45\textwidth]{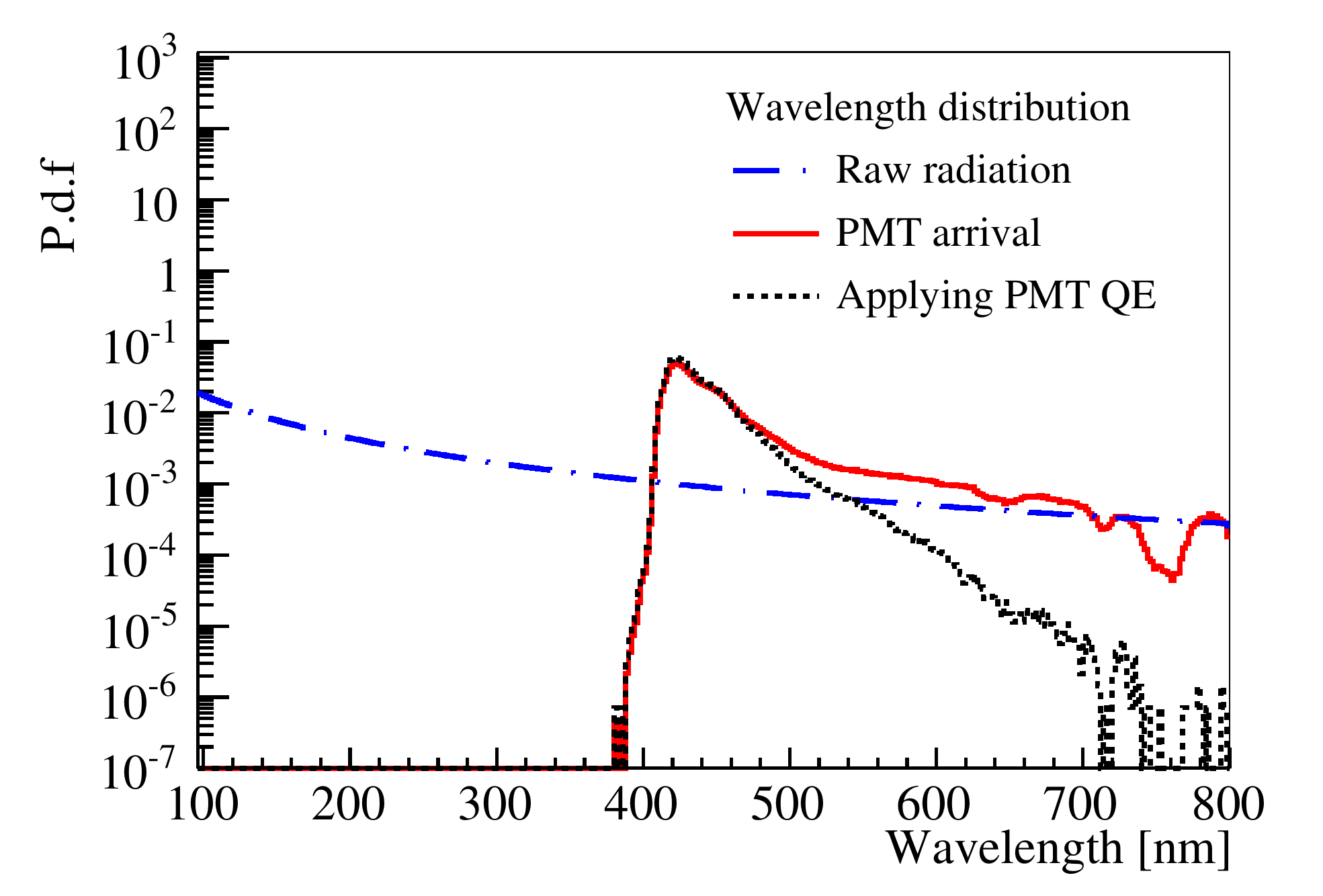}
\caption {\label{fig:CerenkovW} The wavelength p.d.f.~of Cherenkov photons at generation~(blue chain line), the photons arriving at the PMT photocathode~(red solid line), and the ones converted to photoelectrons by the photocathode~(black dashed line). Most Cherenkov photons are absorbed and re-emitted with certain efficiencies dependent on the photon wavelengths.}
\end{centering}
\end{figure}

Cherenkov photons are generated if a particle's velocity is larger than the phase velocity of light in the medium~\cite{Cherenkov}.
The intensity is calculated with the Frank$-$Tamm formula~\cite{FrankTamm}, which is implemented in the simulation of Geant4, generating photons with wavelengths as the blue chain line in Fig.~\ref{fig:CerenkovW}. The spectrum relies on the LS refractive index, thus the imperfect knowledge of the index in the ultra-violet region introduces uncertainties in this.

The LS has good transparency only for photons in a narrow range of wavelengths, from 400~nm to 600~nm.
Photons with other wavelengths would be absorbed by the solvent or solute.
Absorption by the latter can generate another photon (re-emission) with a certain probability.
The red solid line of Fig.~\ref{fig:CerenkovW} shows the wavelengths of photons arriving at the PMT photocathode after several meters propagation. 

\begin{figure}[!htb]
\begin{centering}
\includegraphics[width=.45\textwidth]{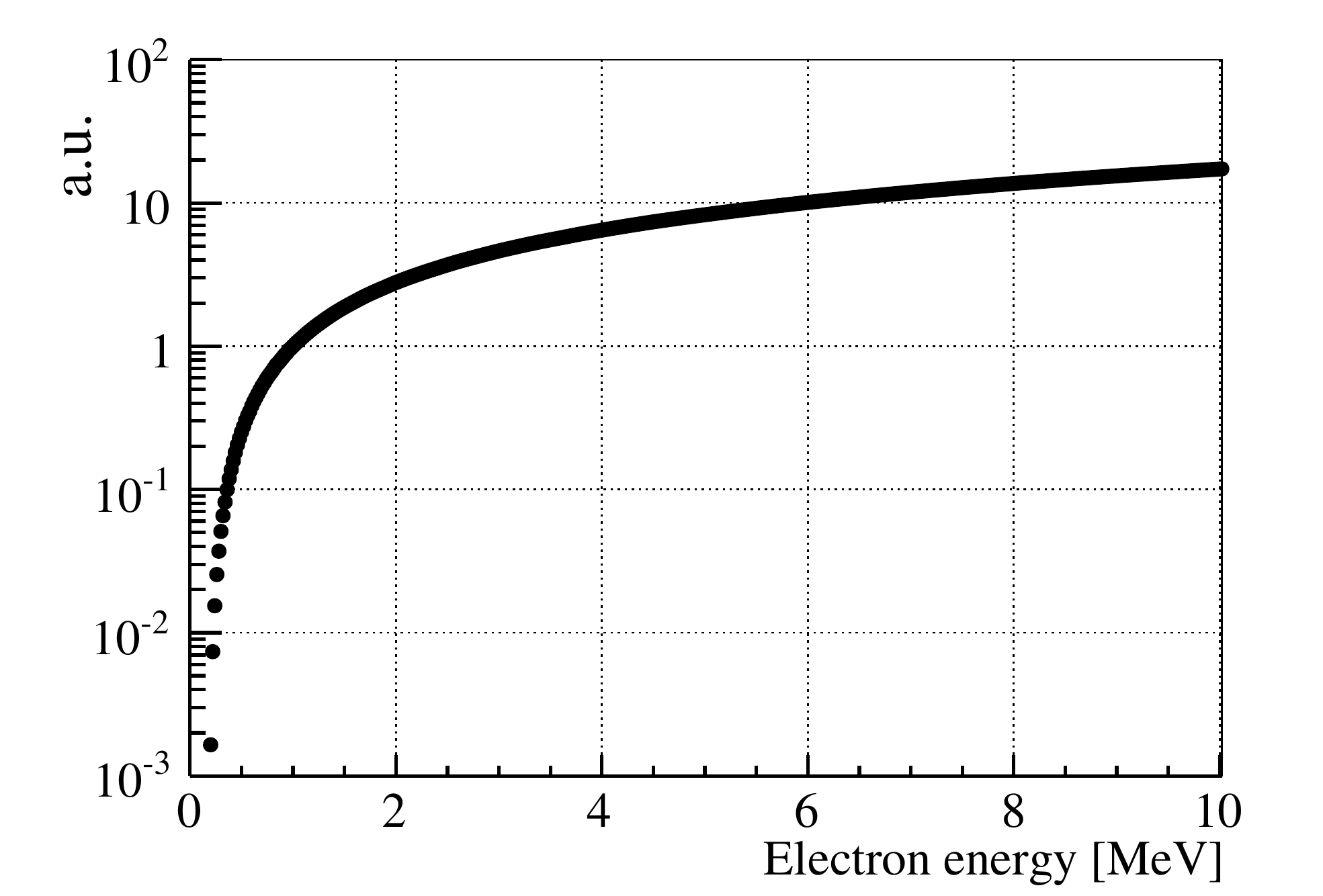}
\caption{\label{fig:Cerenkovradiation}  The Cherenkov contribution to the \electron~kinetic energies, normalized at 1~MeV. The absolute contribution relies on the LS optical properties assumed in the simulation, thus it is a free parameter in the fit described in Sec.~\ref{Results}.}
\end{centering}
\end{figure}

There is no conclusive measurement available for the LS absorption and re-emission,
especially in the ultra-violet and infrared region,
thus the detectable Cherenkov contribution heavily depends on the LS properties assumed by the simulation. Given the fact that the wavelength spectrum of Cherenkov photons has negligible dependence on the primary particle's energy, in this analysis we only use the shape of the Cherenkov contribution to the \electron~energies generated by the Geant4 simulation as shown in Fig.~\ref{fig:Cerenkovradiation},
and determine the normalization by comparing the energy model predictions with the calibration data.

In summary, the scintillation nonlinearity of \electron~can be defined as
\begin{equation}
\label{eq:electronscintillation}
\begin{split}
&f_{\rm ScintNL}^{e-}({\rm E}_{\rm dep},~kB,~k_C,~A) \equiv \frac{{\rm E}_{\rm vis}}{{\rm E}_{\rm dep}} \\
& = A\times\left(f_{q}({\rm E}_{\rm dep}, k_{B}) + k_{C}\times \frac{f_{C}({\rm E}_{\rm dep})}{{\rm E}_{\rm dep}}\right),
\end{split}
\end{equation}
where $f_{q}$~is the quenching curve with a given Birks coefficient $k_{B}$,
$k_{C}$~and $f_{C}$ are the normalization factor and the shape of Cherenkov contribution~(\ref{fig:Cerenkovradiation}), respectively, and $A$~accounts for the absolute energy scale calibration with the neutron capture on gadolinium.

\subsubsection{Scintillation nonlinearity model of the $\gamma$}
\label{gamma}

The $\gamma$ carries zero charge, therefore it cannot directly excite the LS molecules.
Its deposited energy relies on three processes:
the photoelectric effect, Compton scattering, and \positron/\electron~pair production.
In LS Compton scattering is dominant for $\gamma$~energies above about 100~keV; otherwise the photoelectric effect dominates.
Pair production is a secondary process in the energy region of the reactor \nuebar~analysis~(0 to 12~MeV), since the nuclei in LS are light.

\begin{figure}[!htb]
\begin{centering}
\includegraphics[width=.45\textwidth]{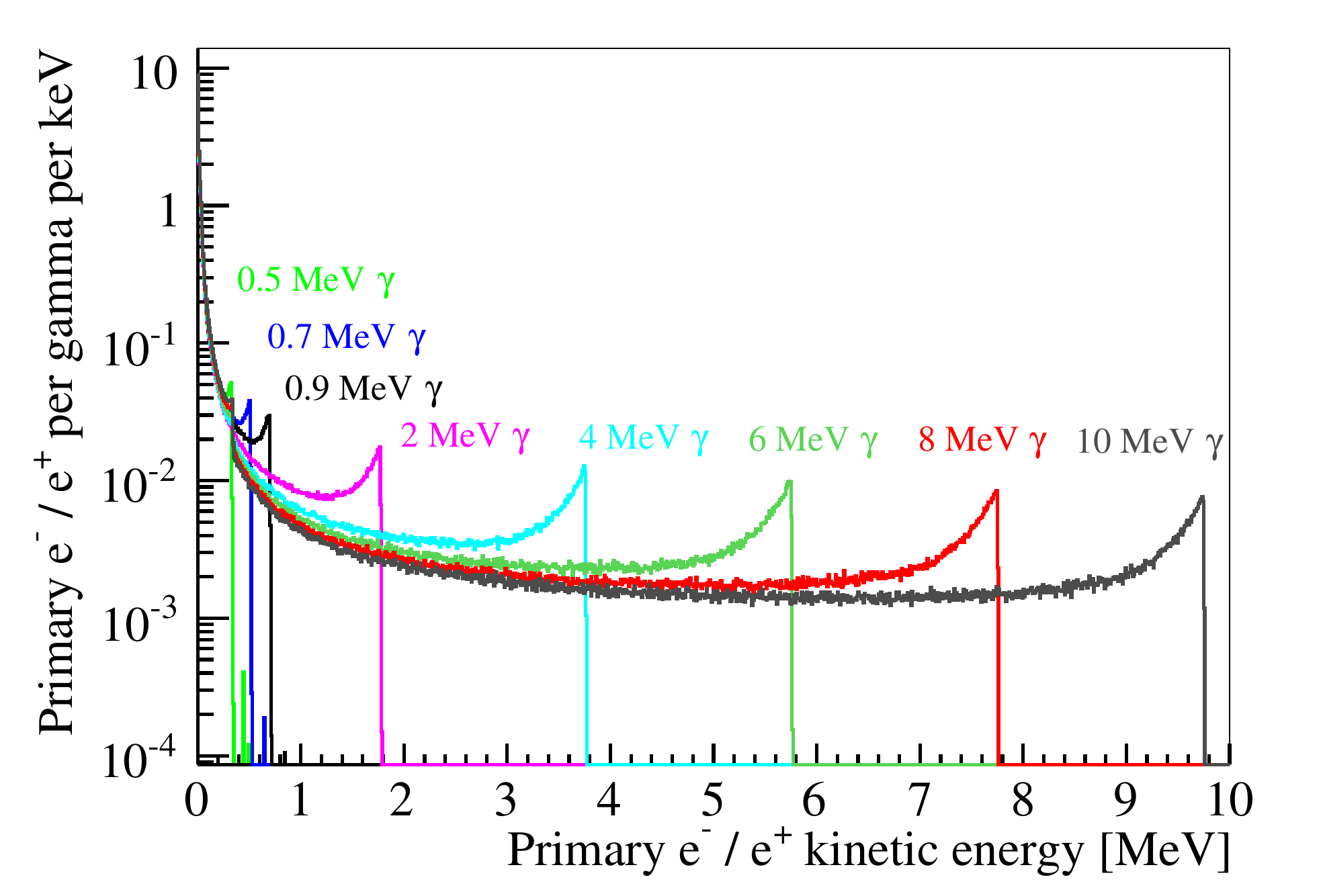}
\caption{
\label{fig:gammatoe} The kinetic energy distributions of induced \elecposi~for a variety of initial $\gamma$~energies.}
\end{centering}
\end{figure}

To determine the scintillation nonlinearity, the $\gamma$s are simulated in LS with Geant4.
The kinetic energies of induced \elecposi~for a range of initial $\gamma$~energies are shown in Fig.~\ref{fig:gammatoe}. The distribution includes \elecposi~generated by the initial $\gamma$ and
the secondary $\gamma$s from the positron annihilation and bremsstrahlung radiation.
The integral of each distribution equals the average total number of \elecposi~generated by each $\gamma$.
Cross validations were done by varying the electromagnetic models to EMPenelope and EMLivermore in Geant4;
the differences are negligible.

The expected \Evis~of a $\gamma$~is expressed as the weighted sum of the \elecposi~visible energies as
\begin{equation}
\label{eq:gammaScintNL}
{\rm E}^\gamma_{\rm vis} =  {\rm E}^\gamma_{\rm dep} \times \frac{ \int_{0}^{{\rm E}^{e^{\pm}}_{\rm max}} P({\rm E}^{e^{\pm}}) \times {\rm E}^{e^{\pm}} \times f_{\rm ScintNL}^{e-}({\rm E}^{e^{\pm}})d{\rm E}^{e^{\pm}}}{\int_{0}^{{\rm E}^{e^{\pm}}_{\rm max}} P({\rm E}^{e^{\pm}}) \times {\rm E}^{e^{\pm}}d{\rm E}^{e^{\pm}}},
\end{equation}
where ${\rm E}^{e^\pm}_{\rm max}$~is the maximum kinetic energy of the induced \elecposi,
$P({\rm E}^{e^{\pm}})$ is the number of \elecposi~with the kinetic energy ${\rm E}^{e^{\pm}}$, and
$f_{\rm ScintNL}^{e-}$~is the \electron~scintillation nonlinearity. The denominator is a normalization accounting for a $O(0.2\%)$~missing energy in the simulation due to the production thresholds.

\subsubsection{Scintillation nonlinearity model of the \positron}
\label{positron}

A positron deposits its kinetic energy in a similar way as \electron, via the ionization and excitation in the LS molecules. Following the calculation method in Ref.~\cite{Berger}, the $dE/dx$~differences between \electron~and \positron~are expected to yield a negligible~($<$0.05\%) effect on scintillation nonlinearity. Thus the difference is safely ignored.
Most of the positrons annihilate at rest, generating two 0.511~MeV $\gamma$s.
The expected deposited energy of \positron~is its kinetic energy ${\rm T}^{e+}$ plus the two $\gamma$s, thus:
\begin{equation}
\label{eq:PtrueE}
{\rm E}_{\rm dep}^{e+} = {\rm T}^{e+}+2\times0.511~{\rm MeV}~.
\end{equation}

There are two phenomena that could introduce a bias to Eq.~\ref{eq:PtrueE}: annihilation in flight and formation of positronium.
A small fraction of \positron~annihilates during flight, generating two $\gamma$s with higher energies.
Simulation suggests that the fraction is about 1\% per MeV of kinetic energy.
If the effect is included, the \positron~nonlinearity would change by less than 0.1\%.
To simplify the nonlinearity model, we ignore the annihilation-in-flight component in the model.

In LS, about half of the \positron~directly annihilates while the other half would form positronium which has two spin states:
singlet~(p-Ps) and triplet~(o-Ps).
The singlet has a life time of about 0.5~ns and decays to two 0.511~MeV $\gamma$s.
The triplet has a life time of about 142~ns in vacuum and decays to three $\gamma$s.
In LS most of the o-Ps decays to p-Ps due to chemical reactions, magnetic effects,
or by positronium interactions with the surrounding electrons.
The measurements suggests the o-Ps's life time is about 3.5~ns in LS~\cite{DCPositronium},
thus about 2\% of o-Ps decay to three $\gamma$s and the others to p-Ps.
The total energy of the three $\gamma$s is 1.022~MeV
but the scintillation nonlinearity is slightly different from the two $\gamma$s case.
The phenomenon introduces a less than 0.05\% variation to the energy nonlinearity of \positron,
which is also safely ignored in this analysis.

In summary, the expected visible energy of \positron~is expressed as
\begin{equation}
 \label{eq:positronscintNL}
 {\rm E}_{\rm vis}^{e+} = {\rm T}^{e+}\times f_{\rm ScintNL}^{e-}({\rm T}^{e+}) + 2\times {\rm E}^{\gamma}_{\rm vis}~(0.511~{\rm MeV}).
\end{equation}

\subsection{Electronics nonlinearity}
\label{ElectronicsNL}

The electronics nonlinearity at Daya Bay is a combined effect from the time distribution of
detected light and the response of the readout electronics. In this section, the origin and the direct measurement of the nonlinearity are described in detail, and an improved energy reconstruction with the electronics nonlinearity corrected at the single channel level is introduced.

\subsubsection{Origin of the electronics nonlinearity}
\label{ElectronicsNLOrig}

As mentioned in Sec.~\ref{DayaBay}, the PMT signal is shaped by a CR-(RC)$^4$ circuit. The time constant of each RC and CR circuit is 25~ns, which means the effective charge integral window is about 100~ns. PMT pulses within a 100~ns time window will form one peak after shaping, and the peak value is used for the charge estimate of these pulses. This charge reconstruction method works well if there is only one pulse in the readout window, or multiple pulses being fully overlapped, otherwise the charge is underestimated.

\begin{figure}[!htb]
\begin{centering}
\includegraphics[width=.50\textwidth]{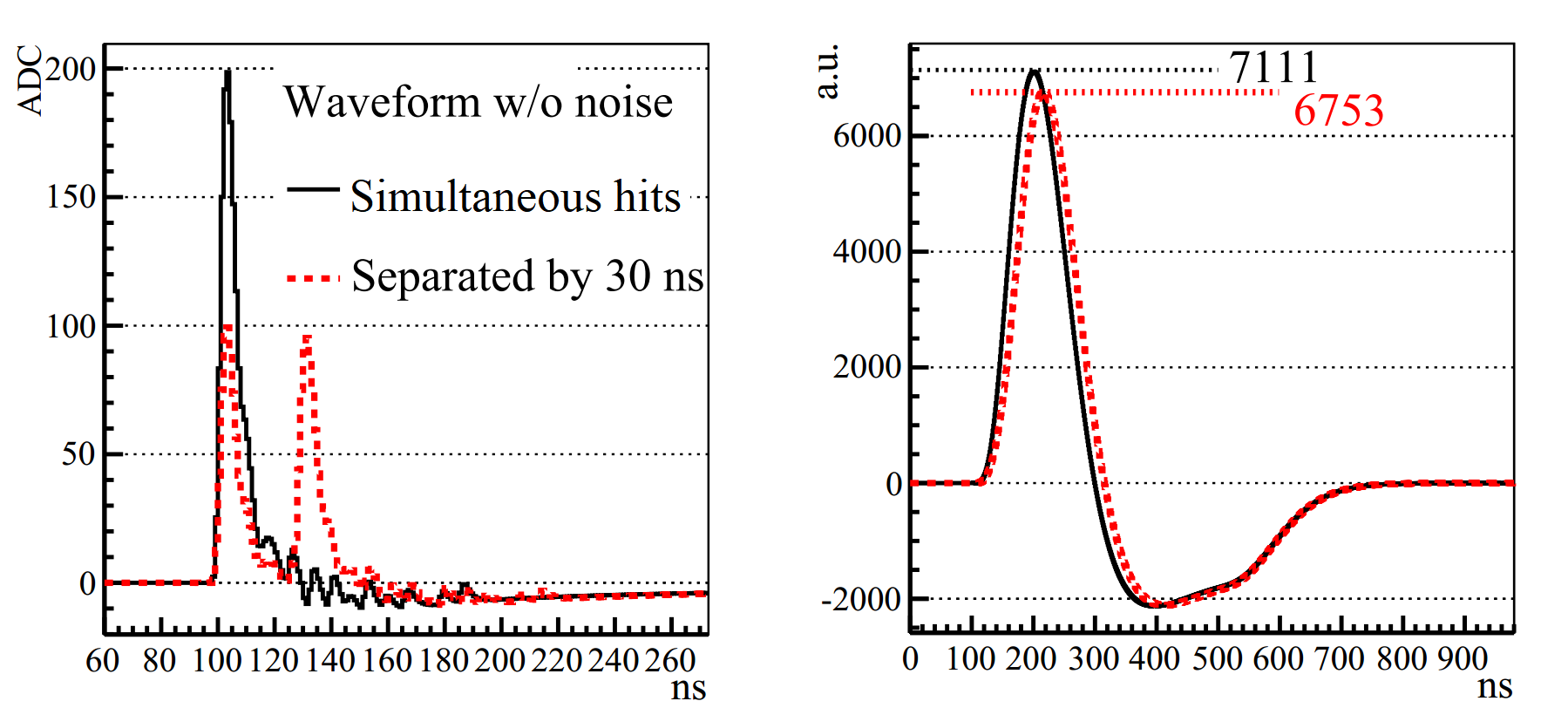}
\caption{\label{fig:CRRC} Left: simulated PMT waveforms based on the FADC measurements, with two simultaneous pulses~(black solid line) and two equal pulses separated by 30~ns~(red dashed line).
The total charges of the two waveforms are the same. Right: waveforms after the CR-(RC)$^4$ shaping circuit in the Daya Bay FEE. The peak values are used to reconstruct the total charges, which is underestimated for the case with two separated pulses~\cite{DYBFADC}.}
\end{centering}
\end{figure}

Figure~\ref{fig:CRRC} shows two simulated raw PMT waveforms~(left) and the ones after shaping~(right). One waveform consists of two simultaneous pulses, and the other one has two pulses separated by 30~ns. The total true charges of the two waveforms are the same.
The reconstructed charge is obtained with the peak value of the waveform after shaping, as 7111 and 6753 in the right plot. For the example with separated pulses, the reconstructed charge is underestimated by 5\%, and larger time separations lead to even larger underestimates of the charge~\cite{DYBFADC}.
Since the LS of Daya Bay is found to have three timing components, with decay constants of $\sim$5~ns, $\sim$30~ns, and $\sim$150~ns, the interplay between the LS timing and the electronics response introduces a 10\% nonlinearity in the energy region of interest~(0 to 12~MeV).

\subsubsection{Direct measurement of the electronics nonlinearity}
\label{ElectronicsNLMeasurement}

To directly measure the electronics nonlinearity, a full FADC readout system was installed on EH1-AD1 in Dec.~2015 and acquired data simultaneously with the FEE readout system. PMT signals were split, then sent to the FEE and FADC system after 1 and 4 times amplification, respectively. The two systems shared the same trigger signal, and the FADC recorded PMT waveforms in a 1-$\mu$s readout window at 1~GHz with a 10-bit resolution. Details of the FADC system can be found in Ref.~\cite{DYBFADC}.

\begin{figure}[!htb]
\begin{centering}
\includegraphics[width=.45\textwidth]{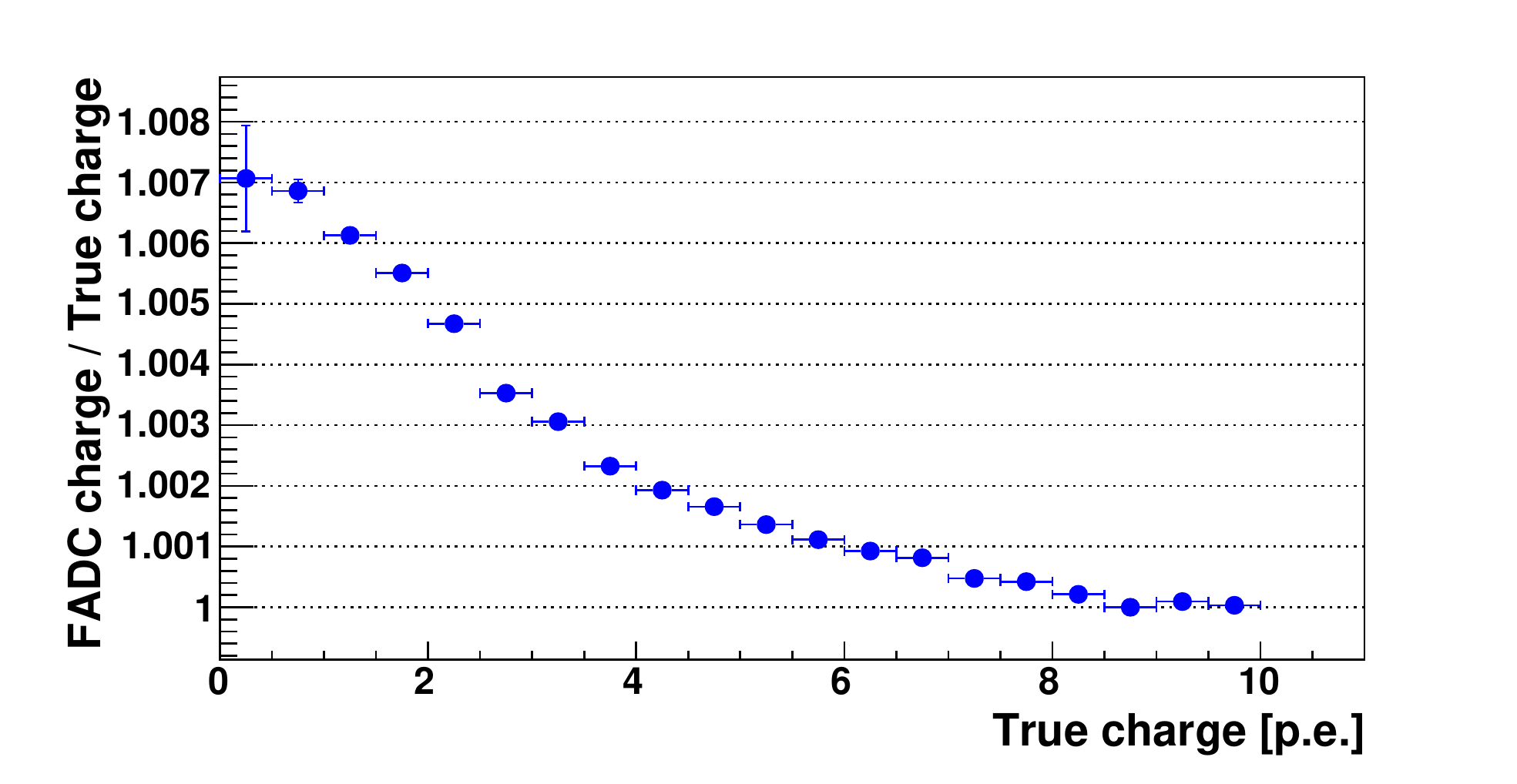}
\caption{\label{fig:SCResidualNL} Residual nonlinearity in the FADC waveform reconstruction of a single channel, estimated with an electronics simulation tuned to data.}
\end{centering}
\end{figure}

The difficulty in the FADC waveform charge reconstruction comes from the PMT signal overshoot which is introduced by the AC circuit and has a $\sim$10\% amplitude to the main pulse, as shown in Fig.~\ref{fig:CRRC}.
The simple waveform integral would result in a 10\% nonlinearity in the estimated charge.
A deconvolution method is used to minimize the dependence on the pulse shape of a single photoelectron,
and to extract the integrated charge with minimum bias.
An electronics simulation is developed and is tuned to have a better than 1\% agreement with data.
The nonlinearity in the reconstructed charge is estimated to be less than 1\% with the simulation, as shown in Fig.~\ref{fig:SCResidualNL}. Details of the waveform reconstruction and simulation validation are also reported in Ref.~\cite{DYBFADC}.

\begin{figure}[!htb]
\begin{centering}
\includegraphics[width=.45\textwidth]{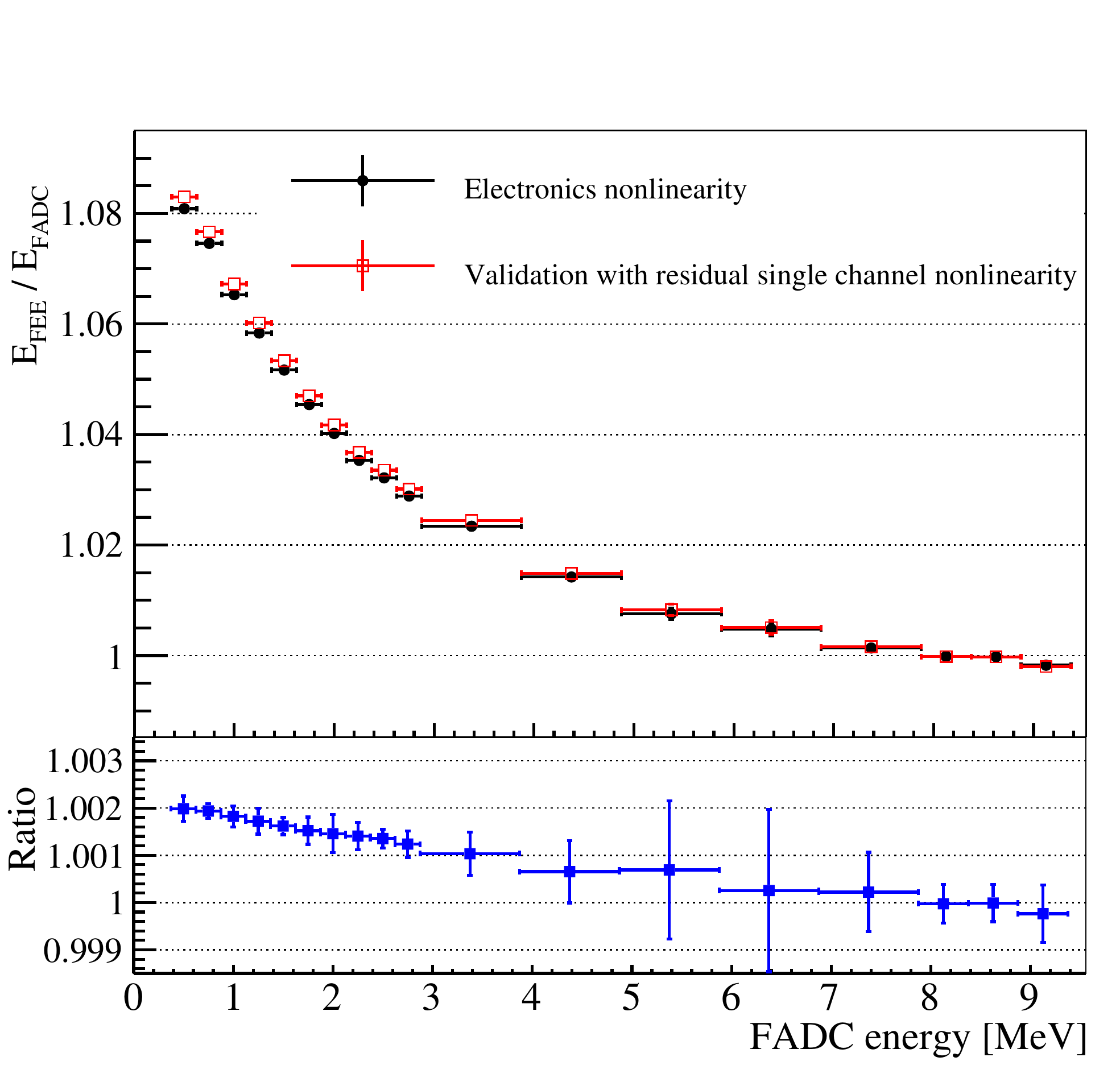}
\caption{\label{fig:fullDetectorElecNL} Upper panel: the measured electronics nonlinearity~(black dots) in the Gd-LS volume, defined as the ratio of the energy given by the FEE system~(E$_{\rm FEE}$) to the one given by FADC system~(E$_{\rm FADC}$). The red squares represent the variation if the simulated residual nonlinearity in the FADC waveform charge reconstruction in Fig.~\ref{fig:SCResidualNL} is propagated to the full detector measurement. The bottom panel shows the comparison between them.}
\end{centering}
\end{figure}

To compare the energies measured by the two electronics systems, similar energy reconstruction strategies are applied, including the single p.e.~calibration, energy scale calibration with neutron capture on gadolinium and the nonuniformity correction. Figure~\ref{fig:fullDetectorElecNL} shows the measured nonlinearity of the FEE readout, which is defined as the ratio of reconstructed energies given by the FEE and FADC readouts. A similar exponential shape is also found in the Daya Bay simulation. The electronics nonlinearity is parameterized as
\begin{equation}
 \label{eq:elecs}
\begin{split}
&f_{\rm ElecNL}({\rm E}_{\rm vis},~\alpha,~\tau) \equiv \frac{{\rm E}_{\rm rec}}{{\rm E}_{\rm vis}} \\
& = 1 + \alpha \times \left(e^{-\frac{{\rm E}_{\rm vis}}{\tau}} - e^{-\frac{8.05~{\rm MeV}}{\tau}}\right),
\end{split}
\end{equation}
where $\alpha$~determines the amplitude of the nonlinearity, and $\tau$~sets the energy dependence. The electronics nonlinearity is set to 1 at the reference energy~(8.05~MeV) of neutron capture on gadolinium. With the parameterized electronics nonlinearity, the full nonlinearity model of \elecposi~and $\gamma$~can be expressed as
\begin{equation}
 \label{eq:fullNL}
 \begin{split}
 &f_{\rm FullNL}({\rm E}_{\rm dep},~A,~kB,~k_C,~\alpha,~\tau) = \frac{{\rm E}_{\rm rec}}{{\rm E}_{\rm dep}} \\
 & = \frac{{\rm E}_{\rm vis}}{{\rm E}_{\rm dep}} \times f_{\rm ElecNL}({\rm E}_{\rm vis})~,
 \end{split}
\end{equation}
where the \Evis~of \electron, \positron, and $\gamma$~are given in Eq.~\ref{eq:electronscintillation}, Eq.~\ref{eq:positronscintNL}, and Eq.~\ref{eq:gammaScintNL}, respectively.

To propagate the residual nonlinearity in the FADC's single channel charge reconstruction~(Fig.~\ref{fig:SCResidualNL}) to the detector level, the reconstructed charge of each FADC channel is corrected with this curve, then the energy reconstruction strategy is applied. The comparison of the reconstructed energies between the corrected FADC and FEE results in an about 0.2\% variation shown as the red squares in Fig.~\ref{fig:fullDetectorElecNL}. The variation is assumed as the systematic uncertainty of the measured electronics nonlinearity.

\subsubsection{Nonlinearity correction at the single channel level}

\begin{figure}[!h]
\begin{centering}
\includegraphics[width=.45\textwidth]{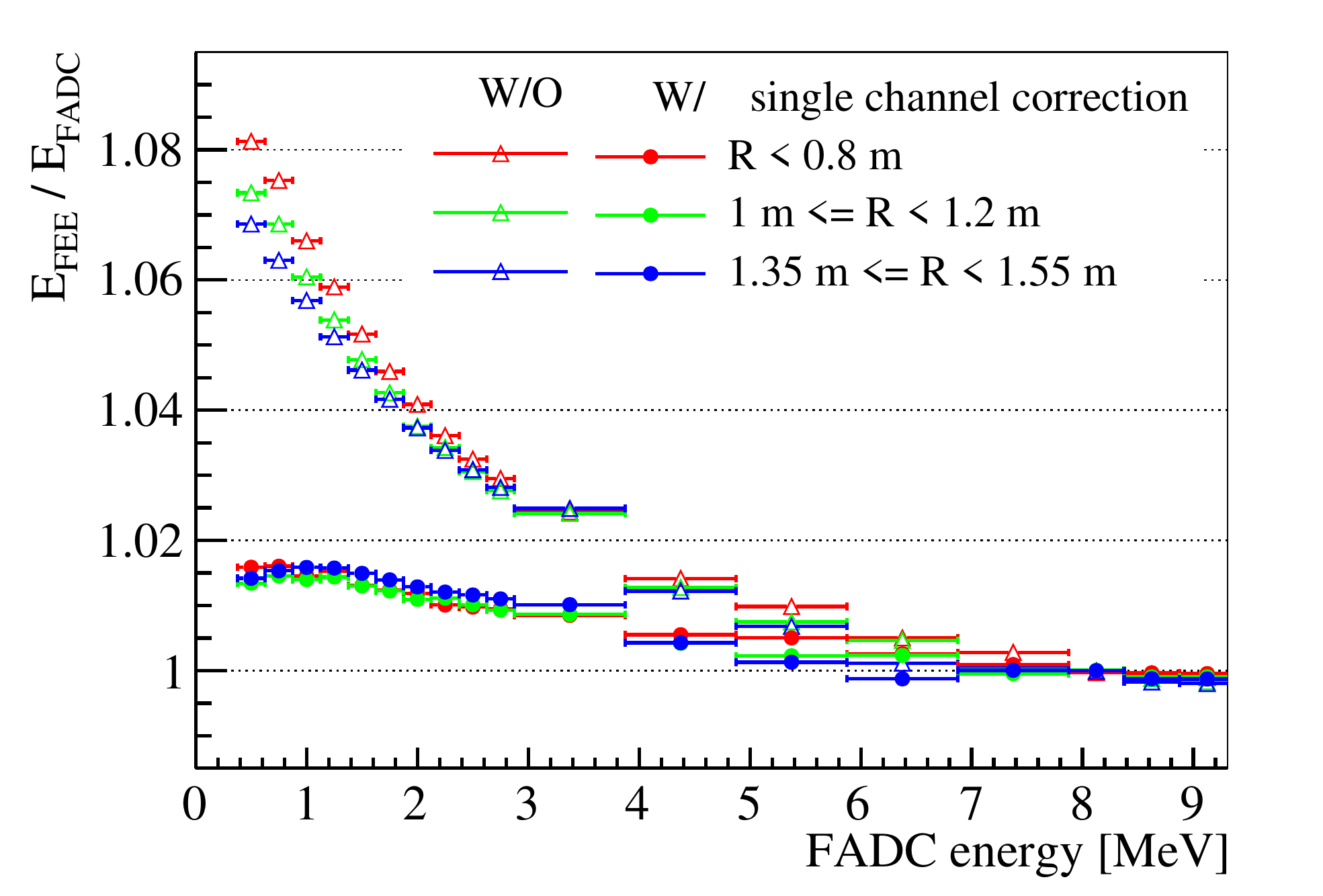}
\caption{\label{fig:electronicsNLcorr} Detector level electronics nonlinearity for the events in three cylinder or hollow cylinder volumes, without~(triangles) and with~(dots) the correction at channel level.  The majority of the position dependence is removed with the correction. See the text for detailed discussions. }
\end{centering}
\end{figure}

The PMT charge pattern varies with the position of a particle interaction within the detector, primarily from the geometric acceptance of the PMTs. Convolving the nonlinearity in the FEE charge reconstruction, a position dependence of the detector level electronics nonlinearity is expected. In data the electronics nonlinearity of events in three cylinder or hollow cylinder volumes with different radiuses and a 3-meters height is compared, as the triangles in Fig.~\ref{fig:electronicsNLcorr}. An about 1\% position dependence is found for events with energies less than 1.5~MeV. The behavior is well reproduced in the simulation.

\begin{figure}[!h]
\begin{centering}
\includegraphics[width=.45\textwidth]{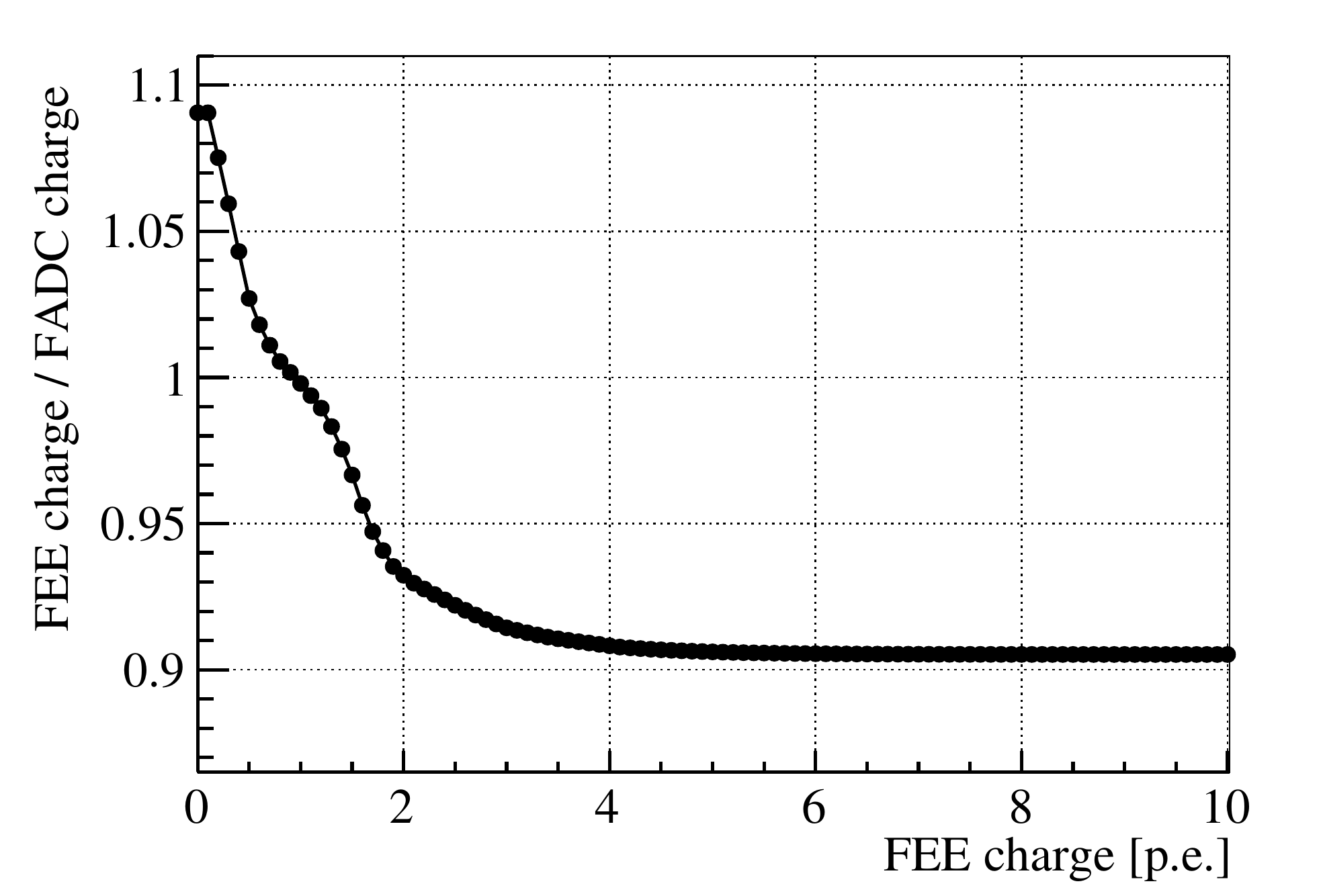}
\caption{\label{fig:singleC} The correction for the single channel nonlinearity, which is defined as the ratio of the charge reconstructed by FEE to the one reconstructed by FADC. The correction is derived from the average of the 192 channels of EH1-AD1. }
\end{centering}
\end{figure}

To remove this effect, the FEE nonlinearity is further corrected at the channel level, yielding a new reconstructed energy. The variation in the nonlinearity of the 192 channels of EH1-AD1 is less than 1\%, and a correction curve is based on the average nonlinearity of them, as shown in Fig.~\ref{fig:singleC}. The curve is assumed to be the same for all the channels of all ADs. The energy reconstruction is modified as
\begin{equation} \label{eq:recENL}
  {\rm E}_{\rm rec} = \left(\sum_i \frac{F_{\rm elec}({\rm Q}_{i})}{\overline{{\rm Q}}^{\rm SPE}_{i}(t)}\right) \cdot \frac{f_{\rm act}(t)}{N'_{\rm PE}(t)} \cdot f'_{\rm pos}(\mathbf{r}_{\rm rec}, t),
\end{equation}
where $F_{\rm elec}$ is the correction shown in Fig.~\ref{fig:singleC}.
The detector energy scale, $N'_{\rm PE}$, increases by about 8\% compared to the one without correction.
The nonuniformity correction, $f'_{\rm pos}$, has the same functional form as the one in Eq.~\ref{eq:recE} but with different coefficient values.
After the channel level correction, the new reconstructed energy is also compared to the FADC energy as shown in Fig.~\ref{fig:electronicsNLcorr}. More than 80\% of the detector level electronics nonlinearity is removed, as well as majority of the position dependence.

In the following the reconstructed energy is without the single channel correction unless otherwise specified, because it was used in the published Daya Bay results~\cite{DYBnGd2016,DYBSpec2017,DYBOsc2018}. The best-fit nonlinearity model of the energy with the single channel correction will also be given in Sec.~\ref{Results}.

\section{Inputs of the energy nonlinearity calibration}
\label{Calibration}

At Daya Bay, ten $\gamma$~calibration points, a continuous $\beta$~spectrum from \B~decays, and the measured electronics nonlinearity are utilized to determine the energy nonlinearity model. The electronics nonlinearity has been described in Sec.~\ref{ElectronicsNL}. This section discusses the measurements, predictions, and the systematic uncertainties of the others.

\subsection{The $\gamma$~calibration points}
\label{GammaData}

The $\gamma$s come from the deployed $^{68}$Ge, $^{137}$Cs, $^{54}$Mn, $^{60}$Co, $^{40}$K
sources, and the naturally occurring \Tl~from the surrounding materials.
For the deployed sources, the accompanying $\beta$~particle is absorbed by the source enclosure.
There are also several $\gamma$s from the three neutron calibration sources based on the~($\alpha,~n$)~reaction of Pu-$^{13}$C, $^{241}$Am-$^{13}$C, and $^{241}$Am-$^{9}$Be. The $\gamma$~energies and the enclosure materials are summarized in Table~\ref{table:gammaData}.

\subsubsection{The deployed $\gamma$~calibration sources}

During physics data taking,
the $^{68}$Ge, $^{60}$Co, and $^{241}$Am-$^{13}$C sources are stored in the ACUs
and deployed weekly into the detector.
The other sources were utilized during special calibration campaigns in the summer of 2012 and January 2017. The detector energy scale is calibrated with the $^{60}$Co source before and
after the deployment of any other sources,
resulting in an energy scale stability better than 0.1\%.

\begin{figure}[!htb]
\begin{centering}
\includegraphics[width=.45\textwidth]{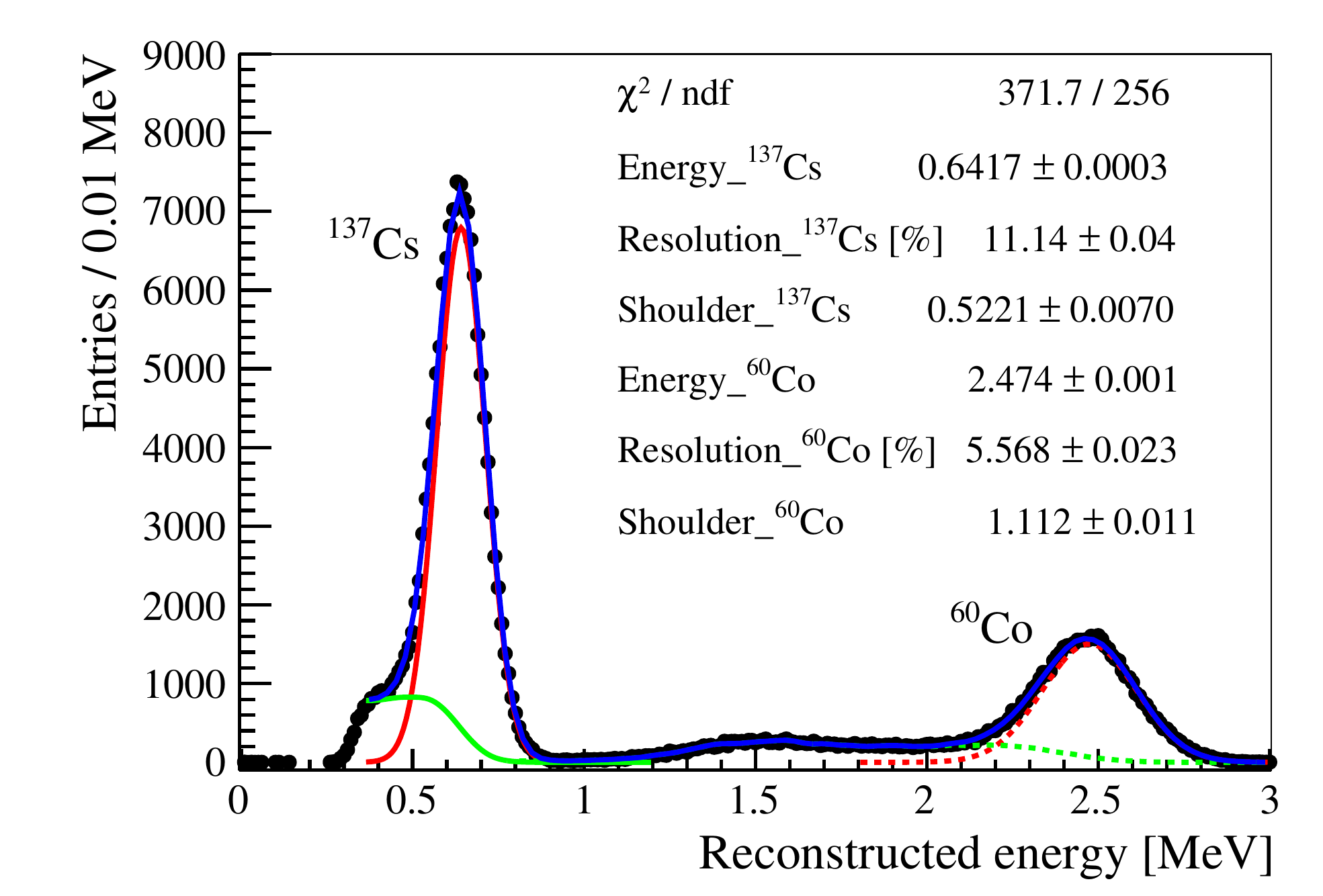}
\caption{\label{fig:GammaFitting} An example of the fit to the $^{137}$Cs~(left peak) and $^{60}$Co~(right peak) combined source. The shoulders in green are due to the $\gamma$~energy loss in the source enclosure. The full energy peaks in red are modeled with Gaussian functions. }
\end{centering}
\end{figure}

The measured spectra of the deployed $\gamma$~calibration sources are obtained by subtracting the spectra taken in the immediately preceding and following runs without the source.
Rejections of events from PMT flashers and muon vetos follow the same methods used in the \nuebar~analysis~\cite{DYBnGd2016}.
The spectrum of a combined source of $^{137}$Cs and $^{60}$Co is shown in Fig.~\ref{fig:GammaFitting}.
The shoulders~(green) to the left from the peaks~(red) are from the $\gamma$~energy loss in the source enclosure and they are well reproduced in the simulation.
In the fit, the full energy peak is modeled with a Gaussian function
and the shoulder is described by the simulated shape with a free normalization.
The simulation shows that such a procedure reduces the bias of the peak energy to less than 0.2\%.
If the spectrum is fit with empirical functions, which cannot describe the shoulder in a correct way, 
the peak could have bias at a level of percent.

\subsubsection{The deployed neutron calibration sources}

The $\gamma$s from the neutron calibration sources are accompanied by neutrons, forming prompt and delayed signal coincidence pairs, which can be used to select the $\gamma$. The $^{241}$Am-$^{9}$Be source with the reaction of $^9$Be($\alpha,~n$)$^{12}$C, produces a 4.44-MeV $\gamma$~from de-excitation of the first excited state of $^{12}$C, while the $^{241}$Am-$^{13}$C and Pu-$^{13}$C sources with the reaction $^{13}$C($\alpha,~n$)$^{16}$O, can produce a 6.13-MeV $\gamma$~from the decay of the second excited state of $^{16}$O.  In the case of the $^{241}$Am-$^{13}$C source a 1-$\mu$m gold foil was placed between the $^{241}$Am and $^{13}$C to absorb enough of the $\alpha$~energy to inhibit the production of $\gamma$s from the $^{16}$O excited states because the source is stored in the ACU during physics data taking~\cite{DYBAMC}.

The energy deposited by the 4.4-MeV $\gamma$~is contaminated by the deposited energy of recoil protons produced by the energetic~($>$1 MeV) neutrons generated in the~($\alpha,~n$)~reaction from the AmBe source. Similarly, the 4.44-MeV $\gamma$~from the excited $^{12}$C by high energy neutrons, is also contaminated by the neutron scattering on protons before the excitation. For this reason there is not a clean 4.44-MeV $\gamma$~available in the analysis. On the other hand, the neutron accompanying the 6.13-MeV $\gamma$~from the de-excitation of $^{16}$O~has less than 100~keV kinetic energy. Considering the quenching of recoil protons, the correction to \Erec~is only 0.5\%. A 100\% relative uncertainty is assigned to this correction.

The neutron capture on gadolinium releases several $\gamma$s with a total energy of
either 7.94~MeV~(for capture on $^{157}$Gd)~or 8.54~MeV~(on $^{155}$Gd).
There are several models and measurements to describe the $\gamma$~energy spectra, such as the model in Geant4,
the model based on the Nuclear Data Sheet, and measurements by several groups~\cite{DYBSpec2017,GdCaptureMeasure1}.
Large discrepancies are found among these models and measurements.
Detailed discussion can be found in Ref.~\cite{DYBFlux2018}.
In this analysis the gadolinium capture is not used.

\begin{figure}[!htb]
\begin{centering}
\includegraphics[width=.45\textwidth]{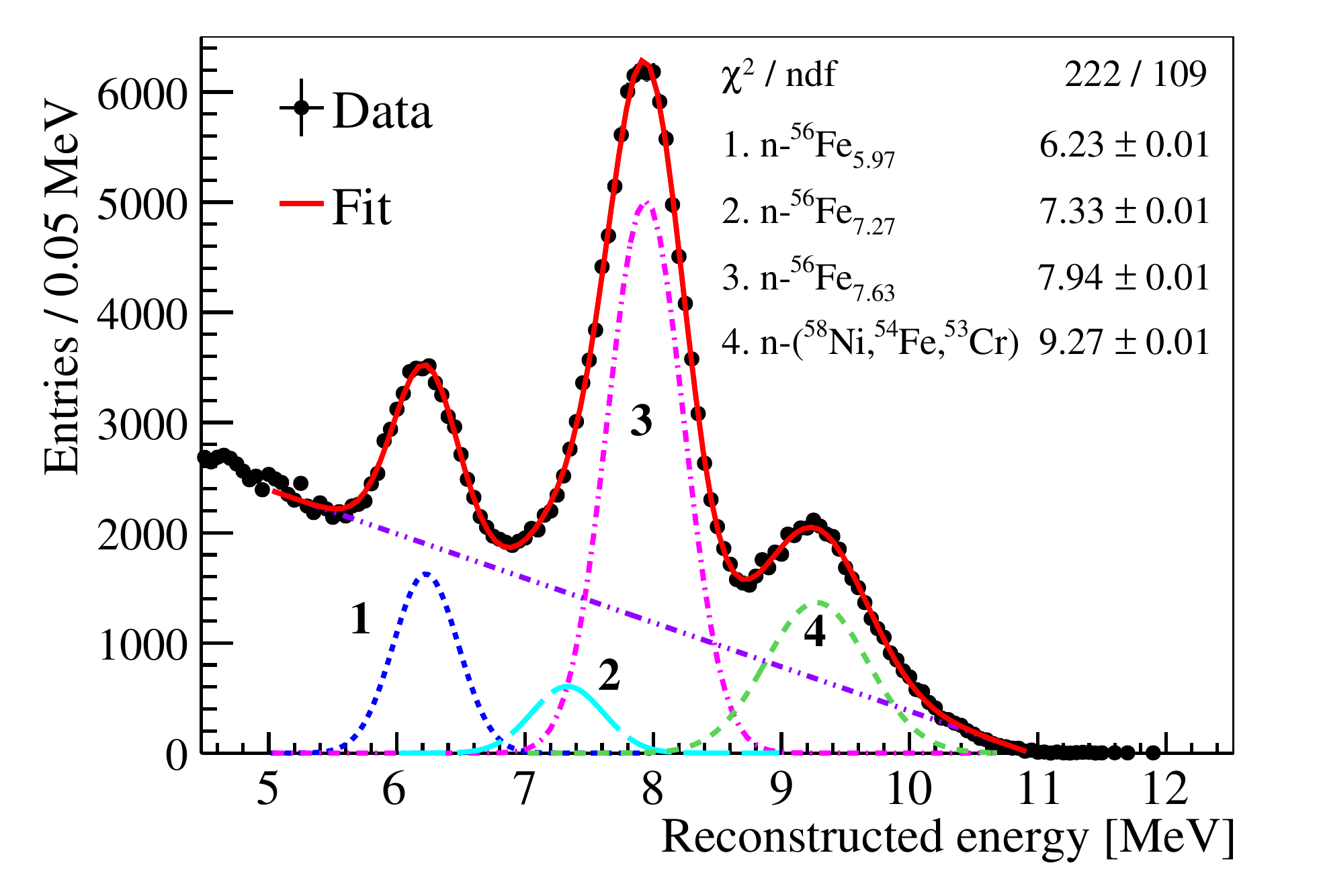}
\caption{\label{fig:nFe} Spectrum of neutron capture on a nucleus in the stainless steel.
The neutrons are from special calibration runs with a $\sim$59~Hz $^{241}$Am-$^{13}$C source placed on the lid of EH3-AD2 in 2012. The subscripted in the legend is the true energy, and the corresponding reconstructed energy is also shown in the unit of MeV. The third peak in this plot is used in this study.}
\end{centering}
\end{figure}

A $^{241}$Am-$^{13}$C source with a neutron emission rate of $\sim$59~s$^{-1}$
was temporarily installed on the lid of EH3-AD2 for more than 10 days in the summer of 2012~\cite{DYBnGd2016}.
The neutron captures on a nucleus~(Fe, Cr, Ni) in the steel are observed, as shown in Fig.~\ref{fig:nFe}. The spectrum is fitted with an empirical background shape, and the third peak of the neutron capture on $^{56}$Fe is used to constrain the energy nonlinearity in the high energy region. 
Since the neutron capture concentrates at the top of the detector,
the residual nonuniformity is examined and found to be less than 0.5\% with different vertex cuts.
A conservative 1\% uncertainty is assigned to the \Erec~of this $\gamma$, after considering the background modeling and residual nonuniformity.

\begin{table*}[tb]
  \caption{The $\gamma$~calibration points used in this analysis. A $'+'$~or $'2\times'$~in column 2 indicates two $\gamma$s are produced by the source. For sources with $\gamma$s with different branching ratios, the ratios are given in parentheses. An optical shadowing~(OS) correction is applied to take into account absorption of optical photons by the source enclosures. The correction size is estimated with simulation and the corrected \Erec~is used in the nonlinearity model fit. Relative uncertainties of corrected \Erec~are uncorrelated among the $\gamma$~calibration points.}
  \label{table:gammaData}
  \small
  \begin{tabular}{c|c|c|c|c|c|c}
    \hline \hline
    Source & Energy of $\gamma$(s) & Enclosure material & \Erec &OS correction& Corrected \Erec& Uncertainty \\
    \hline
    $^{137}$Cs & 0.662~MeV & Teflon + SS &0.642~MeV &0.4\% &0.649~MeV &0.5\% \\ \hline
    $^{54}$Mn & 0.834~MeV & Teflon + SS &0.827~MeV &0.2\% &0.835~MeV &0.5\% \\ \hline
    $^{68}$Ge & 2$\times$0.511~MeV&Acrylic + SS&0.958~MeV &0.75\%  &0.966~MeV &0.5\% \\ \hline
    $^{40}$K &1.461~MeV & Teflon + SS &1.475~MeV &1.2\%&1.493~MeV &0.5\% \\ \hline
    n-H &2.223~MeV & Acrylic + SS & 2.292~MeV &0.36\%&2.300~MeV &0.5\% \\ \hline
    $^{60}$Co & 1.173~+ 1.332~MeV & Acrylic + SS &2.474~MeV &0.65\% &2.490~MeV &0.5\% \\ \hline
    $^{208}$Tl & 2.614~MeV & N/A &2.701~MeV &N/A  &2.701~MeV &0.5\% \\ \hline
    n-$^{12}$C & \tabincell{l}{ 4.945~MeV~(68\%)\\1.261 + 3.683~MeV~(32\%)}&Acrylic + SS&5.118~MeV & $<$0.1\%  &5.119~MeV &0.8\% \\ \hline
    $^{16}$O$^{*}$ &6.130~MeV&Acrylic + SS&6.312~MeV &0.6\% &6.350~MeV &0.8\% \\ \hline
    n-$^{56}$Fe& 7.637~MeV& N/A &7.940~MeV &N/A &7.940~MeV&1.4\% \\ \hline
    \hline
  \end{tabular}
\end{table*}

\subsubsection{Correction for the optical shadowing of source enclosures}

\begin{figure}[!htb]
\begin{centering}
\includegraphics[width=.45\textwidth]{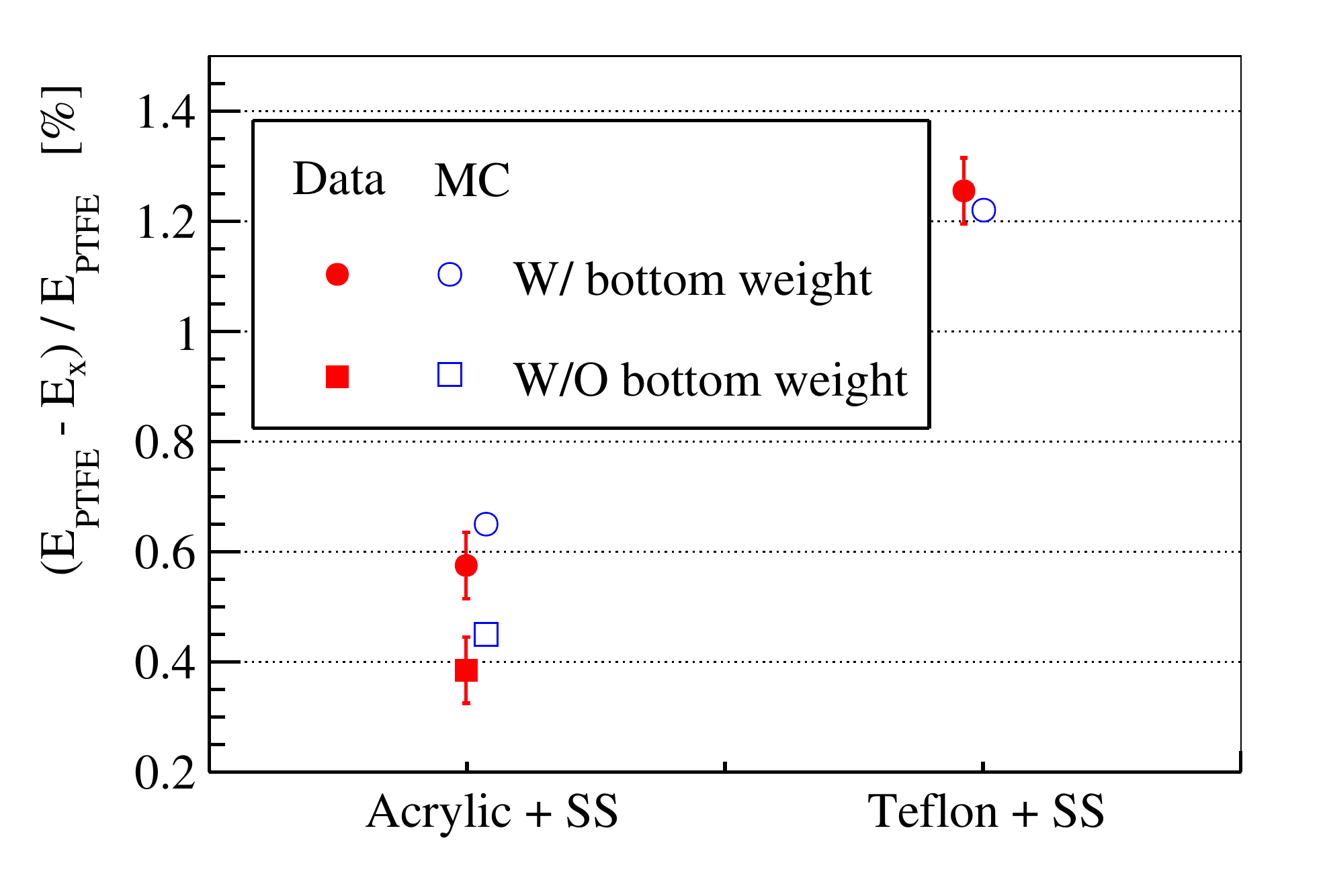}
\caption{\label{fig:OS_diff} Results of special calibration utilizing $^{60}$Co sources stored in a stainless steel chamber and enclosed by PTFE of high reflectivity, acrylic, greenish teflon. A $^{60}$Co source enclosued by acrylic but without the bottom weight is also deployed.}
\end{centering}
\end{figure}

As discussed in Sec.~\ref{DayaBay},
the calibration source enclosures and weights absorb scintillation photons up to 1.5\% depending on the $\gamma$~energy
and the surface reflectivity.
The absorption fractions from the simulation are used to correct the reconstructed energies as listed in Table~\ref{table:gammaData}.
To anchor the simulation, a special calibration was carried out in Jan.~2017.
Several $^{60}$Co sources, stored in a stainless steel chamber and with different enclosures~(PTFE of high reflectivity, acrylic, greenish teflon, acrylic without the bottom weight) were deployed, absorbing $\le$0.1\%, 0.65\%, 1.22\%, and 0.45\% photons, respectively, as determined from the simulation.
Comparison of the data and simulation is shown in Fig.~\ref{fig:OS_diff}, and a better than 0.2\% agreement is found. When varying the electromagnetic models in the Geant4 simulation, a 0.2\% variation is found in the absorption fractions of scintillation photons at different $\gamma$~energies.
Thus a conservative 0.3\% uncertainty is assigned to the correction.

Table~\ref{table:gammaData} summarizes the $\gamma$~calibration points used in this analysis. The uncertainties in the table are the sum of the statistical uncertainty, the optical shadowing correction uncertainty, and the one from spectrum fitting. They are assumed to be uncorrelated among the calibration points.
Large uncertainties are assigned to the $\gamma$s from n-$^{12}$C, $^{16}$O$^{*}$ and n-$^{56}$Fe due to the empirical background model.
The 6.13-MeV $\gamma$ from $^{16}$O$^{*}$ de-excitation has an additional 0.5\% uncertainty because of the
contamination due to prompt neutron scattering.
All the $\gamma$~calibration points except the n-$^{56}$Fe have a correlated 0.3\% uncertainty taking into account
 the residual nonuniformity between point-like $\gamma$~sources,
which preferentially illuminate the detector center, and IBD events over the full target volume.

Given the $\gamma$~energy, the predicted reconstructed energy \Erec~is determined from Eq.~\ref{eq:fullNL}.
If there are multiple $\gamma$s in one event, such as the 1.17-MeV and 1.33-MeV $\gamma$s~of $^{60}$Co ,
\Evis~is the branching ratio weighted sum of visible energies of the multiple $\gamma$s.

\subsection{Continuous $\beta$~spectrum}
\label{B12}
\subsubsection{The measured $\beta$~spectrum from \B~decays}

The average $\beta$~energy and $Q$-value of \B~are about 6.4~MeV and 13.4~MeV, respectively.
The $\beta$~spectrum provides a direct constraint on the nonlinearity model of \electron, especially in the high energy region. An enhanced \B~sample is selected in a~(5~ms, 50~ms)~time window after a showering muon,
while also requiring a reconstructed vertex in the Gd-LS volume.
Showering muons are selected by AD events with \Erec~larger than 2.5 GeV.
The background is estimated with a~(505~ms, 550~ms)~window after the muon.
The start time of the signal window is chosen to eliminate the influence of spallation neutron capture on $^1$H,
and the end time of the signal window is based on the 20.2~ms half-life of \B.

\begin{figure}[!htb]
\begin{centering}
\includegraphics[width=.45\textwidth]{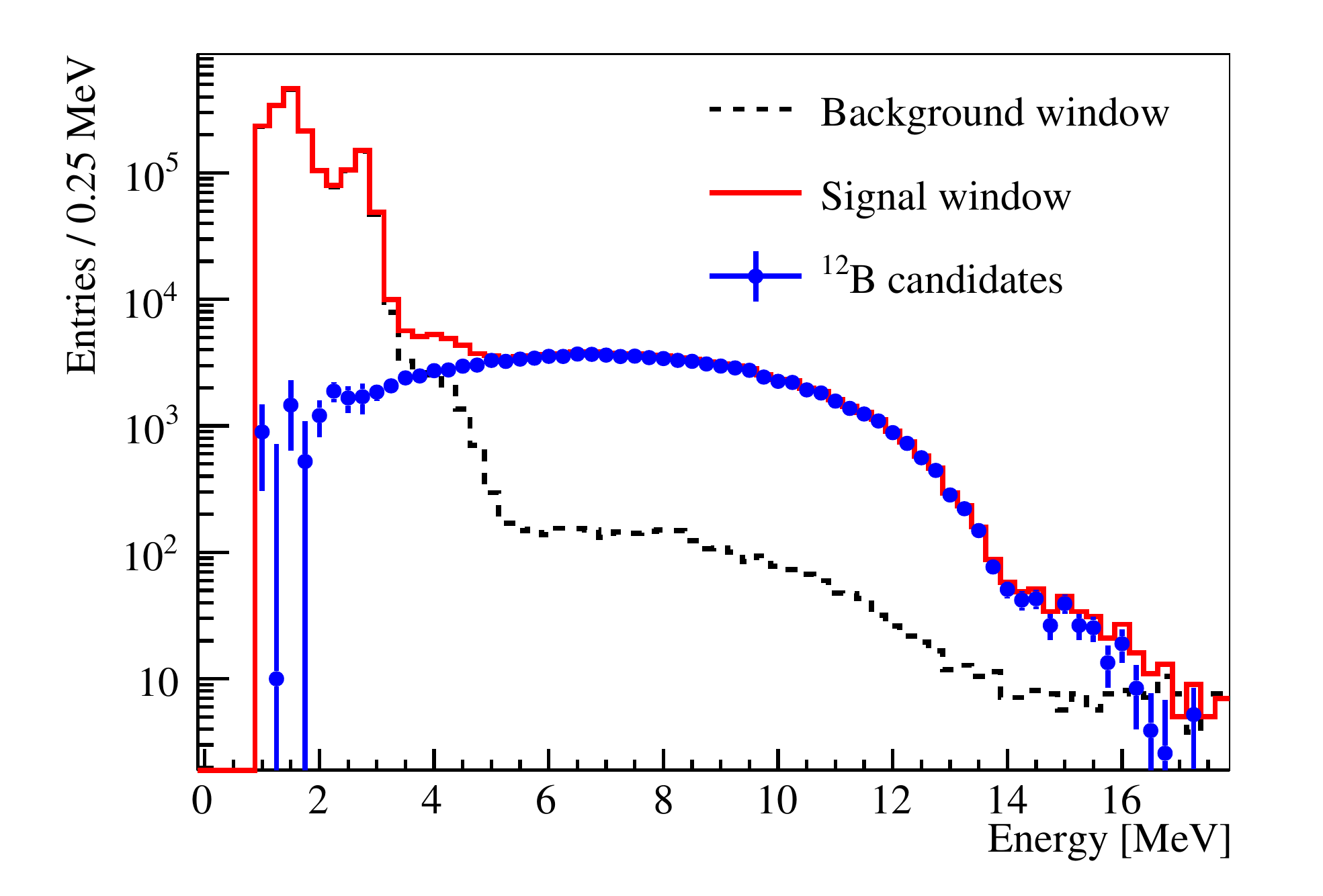}
\caption{
\label{fig:B12_measured} Beta spectrum of $^{12}$B~in EH1-AD1~(blue dots)~obtained by subtracting backgrounds~(black dash line) from an enhanced sample~(red solid line) with the~(5~ms, 50~ms)~window after a showering muon. The background is estimated with the~(505~ms, 550~ms) window after the muon.
}
\end{centering}
\end{figure}

Figure~\ref{fig:B12_measured} shows the \B~signal and background spectra in EH1-AD1.
The event excess in 14~MeV to 18~MeV is from \N~decay, which is a $\beta^+$~isotope with a 11~ms half-life and a $Q$-value of 17.3~MeV. 
The contribution from \N~is estimated to be $(2.7\pm0.8)$\% and is included in the fitting of the energy nonlinearity model.
In the four ADs of EH1 and EH2, a total of about 470,000 events are observed in the 3.5~MeV to 18~MeV energy range
and are used in this analysis. The measured spectrum is provided in the Supplemental Material~\cite{Supplemental}.

\subsubsection{Prediction of the \B~$\beta$~spectrum }

Table~\ref{table:B12Decay} shows the primary decay channels of \B~\cite{B12Decay}.
The first channel is to the ground state of $^{12}$C,
the second to the first excited state of $^{12}$C*~(4.44) which releases a 4.44-MeV $\gamma$,
and the third to the second excited state of $^{12}$C*(7.65) which converts to three $\alpha$~particles.
The other channels, with less than 0.1\% contribution, are safely ignored.
The intensity of the second channel is precisely measured with the $\beta-\gamma$ coincidence,
and that of the third channel is measured by detecting the unbound $\alpha$s.
The intensity of the first channel is obtained by subtracting all the other observed decay channels from unity.
Some measurements suggest the third channel has a branching ratio larger than 0.54\%.
We inflate the uncertainty of this branching ratio to 1\%, to cover these discrepancies.

\begin{table}[tb]
  \caption{Primary decay channels of \B~($J^\pi = 1^+$)~\cite{B12Decay}.}
  \label{table:B12Decay}
  \small
  \begin{tabular}{cccc}
    \\
    \hline
	\tabincell{l}{ Energy of $^{12}$C \\ excited states} & Branching ratio~(\%)  & $J^\pi$ & E$_{{\mathrm{max}}}^\beta$\\
    \hline
    Ground state & 98.300 $\pm$0.028  & $0^+$ & 13.37~MeV \\
	\hline
    4.44~MeV &1.182$\pm$0.019  & $2^+$ & 8.93~MeV \\
	\hline
	7.65~MeV &\tabincell{l}{ 0.54$\pm$0.02 \\ 1.3$\pm$0.4 \\1.1$\pm$0.3 }& $0^+$ & 5.71~MeV \\

    \hline
  \end{tabular}
\end{table}

\begin{figure}[!htb]
\begin{centering}
\includegraphics[width=.45\textwidth]{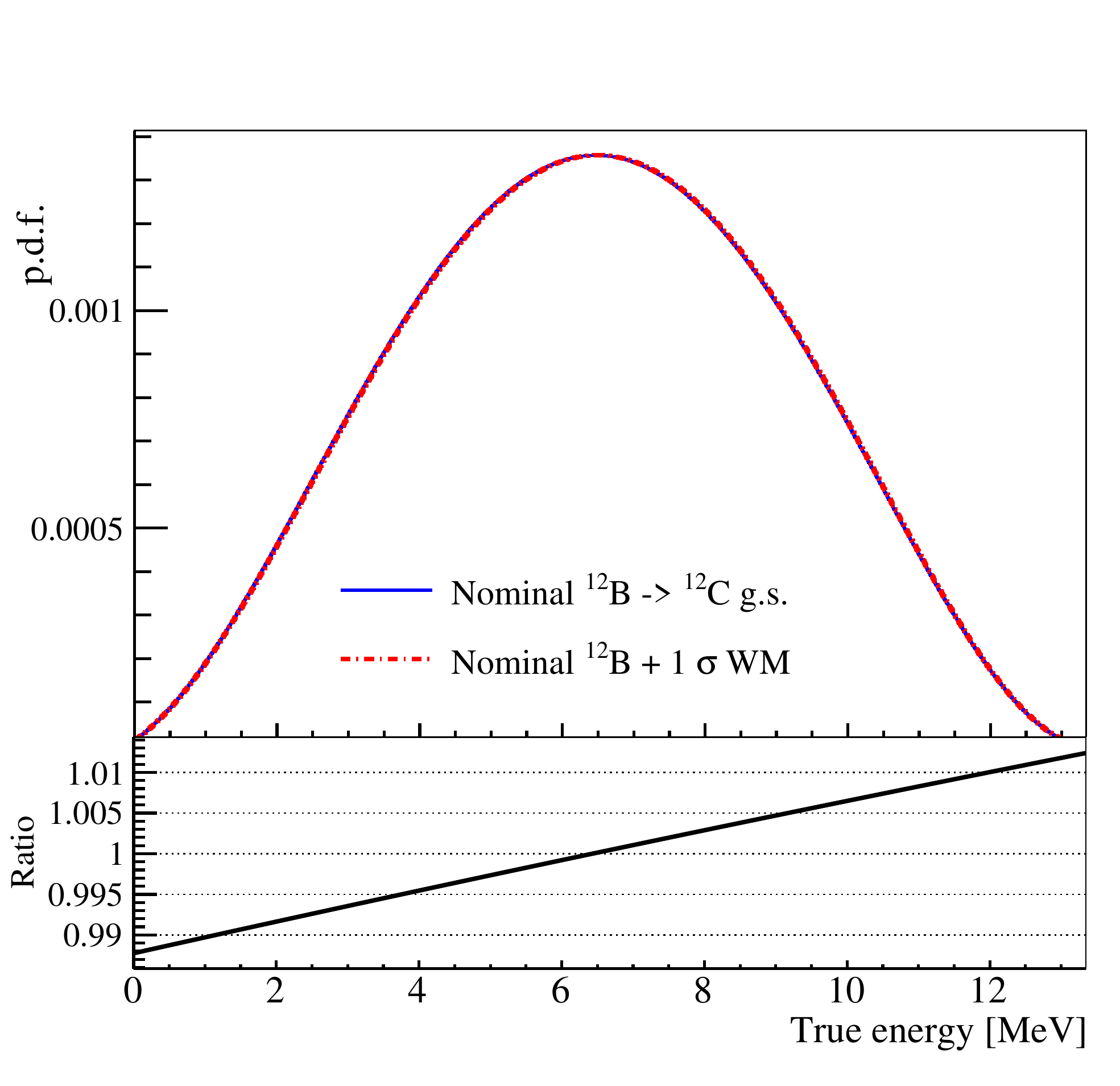}
\caption{\label{fig:B12Prediction} Top panel: Predicted $\beta$~spectrum of \B~decaying to the ground state of $^{12}C$~(blue solid line, nearly obscured by the red dashed line),
and the one if the weak magnetism~(WM) correction is varied by 1$\sigma$~(red dashed line).
Bottom panel: The ratio of the two predictions. }
\end{centering}
\end{figure}

The $\beta$~decay of \B~is an allowed transition of the Gamow-Teller type.
The spectrum prediction follows the discussion in Refs.~\cite{B12Spectrum1,B12Spectrum2}.
We include the shape corrections due to the screening effect, the electromagnetic and weak interaction finite-size effect, and the weak magnetism effect.
Since the exchange of the virtual photon does not affect the $\beta$~shape and the emitted real photon is detected in the LS detector, the radiative correction is not included.
The primary shape uncertainty comes from the weak magnetism correction
which is a linear correction with a coefficient of~(0.48$\pm$0.24)~\%/MeV, as shown in Fig.~\ref{fig:B12Prediction}.

The predicted $\beta$~spectrum of each decay channel is calculated in 10~keV bins and provided in the Supplemental Material~\cite{Supplemental}.
The \electron~scintillation nonlinearity $f_{\rm ScintNL}^{e-}$~in Eq.~\ref{eq:electronscintillation} is applied to each energy bin, resulting in a visible energy spectrum.
To the second channel, \Evis~of the 4.44-MeV $\gamma$ is added to the spectrum. To the third channel, 
a 0.5$\pm$0.1~MeV visible energy is added to take into account the subsequent disintegration of $^{12}$C to 3 $\alpha$s.
The electronics nonlinearity is applied to the visible energy spectrum. Then the energy resolution function is convolved. When summing over the three decay channels, the intensity of the first channel is calculated by subtracting branching ratios of the second and third channels from unity.
Finally the summed spectrum is rebinned to 0.25~MeV/bin to compare with data.

Besides \B, the $\beta$~spectra from \Bi,
\BI, and \Tl~could constrain the energy nonlinearity of low energy electrons~(less than 3~MeV). But most of decay channels are first forbidden nonunique transitions. Prediction of the $\beta$~spectra is difficult since the structure of the initial and final nuclear states need to be considered.
Generally the first forbidden nonunique transitions use predictions of the corresponding allowed ones.
The difficulty is to verify how well the simplification works and how to assess uncertainties in the prediction.
Thus these three isotopes are not used in this analysis.

\section{Results of the energy nonlinearity calibration}
\label{Results}

To determine the five free parameters of the energy nonlinearity model in Eq.~\ref{eq:fullNL}, a $\chi^2$~function with nuisance parameters is constructed for a combined fit on the ten $\gamma$~calibration points, the $\beta$~spectrum from \B~decays, and the directly measured electronics nonlinearity. This section will introduce the definition of the $\chi^2$~function, and the fitted results of the nonlinearity model.

\subsection{Statistical methods}

A $\chi^2$~test statistic with five free parameters is defined as
\begin{equation}
 \label{eq:chi2}
\begin{split}
\chi^{2}&(A,~kB,~k_C,~\alpha,~\tau) = \chi^{2}_{\rm stat} + \chi^{2}_{\rm syst}(\bm \nu), \\
\chi^{2}_{\rm stat} & = \sum_{i=1}^{58}\frac{(P^{B}_{i}-M^B_{i})^2}{(\sigma_B^{i})^2} +
\sum_{i=1}^{10}\frac{(P^{\gamma}_{i}\times(1+\nu_{\gamma})-M^{\gamma}_{i})^2}{(\sigma_{\gamma}^{i})^2} \\
&+ \sum_{i=1}^{18}\frac{(P^{\rm elec}_{i}\times (1+\nu_{\rm elec} \times a_i)-M^{\rm elec}_{i})^2}{(\sigma_{\rm elec}^{i})^2}, \\
\chi^{2}_{\rm syst}(\bm \nu) & = \sum_{j=1}^{3}(\nu_{\rm WM}^{j})^{2}
 + \sum_{j=2}^{3}\left(\frac{\nu^{j}_{\rm BR} - \nu^{j~~\rm {Initial}}_{\rm BR}}{\sigma^{j}_{\rm BR}}\right)^{2}\\
 &+ \frac{(\nu_{{\rm N12/B12}}-\nu^{\rm Initial}_{{\rm N12/B12}})^2}{\sigma^{2}_{{\rm N12/B12}}} + \frac{\nu^2_{\gamma}}{\sigma^{2}_{\gamma}}+ \nu_{\rm elec}^{2},
\end{split}
\end{equation}
for comparison of the measurement~($M$) to the prediction~($P$). The measurements consist of:
1) $M^{\gamma}_{i}$~($\sigma_{\gamma}^{i}$), the ten $\gamma$~calibration points and their uncorrelated uncertainties taken from Table~\ref{table:gammaData}, 2) $M^B_{i}$~($\sigma_B^{i}$), the reconstructed energy spectrum of \B~in 58 energy bins and their statistical uncertainties as discussed in Sec.~\ref{B12}, and 3) $M^{\rm elec}_{i}$~($\sigma_{\rm elec}^{i}$), the measured electronics nonlinearity in 18 energy bins and their statistical uncertainties as shown in Fig.~\ref{fig:fullDetectorElecNL}. The predicted values and uncertainties of these quantities are described in Sec.~\ref{GammaData}, Sec.~\ref{B12}, and Eq.~\ref{eq:elecs}, respectively.

The $\chi^{2}_{\rm syst}(\bm \nu)$~penalizes the total $\chi^2$~based on deviations
of the systematic nuisance parameters from their expected values. The nuisance parameters consist of the correlated uncertainty among the $\gamma$~rays~($\nu_{\gamma}$) except for the one from neutron capture on $^{56}$Fe, the uncertainty of \N~contribution in the \B~spectrum~($\nu_{{\rm N12/B12}}$), the uncertainty of the weak magnetism correction in the \B~prediction~($\nu_{\rm WM}^j$), the branching ratio uncertainties of the second and third \B~decay channels~($\nu_{\rm BR}^j$), and the uncertainty of the measured electronics nonlinearity due to the uncertainty of the single channel's charge reconstruction~($\nu_{\rm elec}$). 

Because the branching ratio uncertainty of the \B~first decay channel is fully correlated to those of the second and third channels, it is not considered a nuisance parameter.
The $\nu_{\rm elec}$~is deduced to follow a standard Gaussian distribution by the coefficients $a_i$, which is the 1$\sigma$ variation of the electronics nonlinearity. The $\nu_{\rm WM}^j$~is assumed to be uncorrelated among the three decay channels, and is also deduced to follow a standard Gaussian distribution. Table~\ref{table:nuisanceparameters} summarizes the free and nuisance parameters, including their definitions, initial values and uncertainties, and the values at the best-fit point.

\begin{table*}[tb]
\begin{centering}

  \caption{Summary of nuisance parameters in the study, including the initial values, estimated uncertainties, and the values at the best-fit point. The $\nu^j_{\rm WM}$ and $\nu_{\rm elec}$ have been deduced to follow the standard Gaussian distribution.}
  \label{table:nuisanceparameters}
  \small
  \begin{tabular}{c|c|c|c|c}
    \hline \hline
	Nuisance parameter & Definition & $\nu^{\rm Initial}$ & Estimated uncertainty & Best-fit value\\ \hline
	$\nu_{\gamma}$     &  \tabincell{l}{ Correlated uncertainty among $\gamma$~points \\ accounting for the residual nonuniformity }   & 0 & 0.003 & 3$\times 10^{-4}$\\ \hline
	$\nu_{{\rm N12/B12}}$ & \tabincell{l}{ Ratio of the $^{12}$N contribution \\ in the measured $^{12}$B spectrum} & 0.027 & 0.008& 0.034\\ \hline
	$\nu^j_{\rm WM}$ & \tabincell{l}{ Uncertainty of the weak magnetism \\ correction in $^{12}$B prediction \\ of the three decay channels} & \tabincell{l}{0 \\ 0 \\ 0} & \tabincell{l}{1 \\ 1 \\ 1} & \tabincell{l}{ -1.43 \\ -0.02  \\ -0.02}  \\ \hline
	$\nu^j_{\rm BR}$ &\tabincell{l}{ Uncertainty of the 2$^{\rm nd}$ and 3$^{\rm rd}$ decay \\ channels of $^{12}$B} & \tabincell{l}{ 0.012 \\ 0.005} & \tabincell{l}{ 2.2$\times 10^{-4}$ \\ 0.01} & \tabincell{l}{ 0.012 \\ 0.017} \\ \hline
	$\nu_{\rm elec}$ &Uncertainty of the measured electronics NL & 0 & 1 & -0.193\\ \hline\hline
  \end{tabular}
  \end{centering}

\end{table*}

\subsection{Results of the nonlinearity model calibration}

\begin{figure}[!h]
\begin{centering}
\includegraphics[width=.45\textwidth]{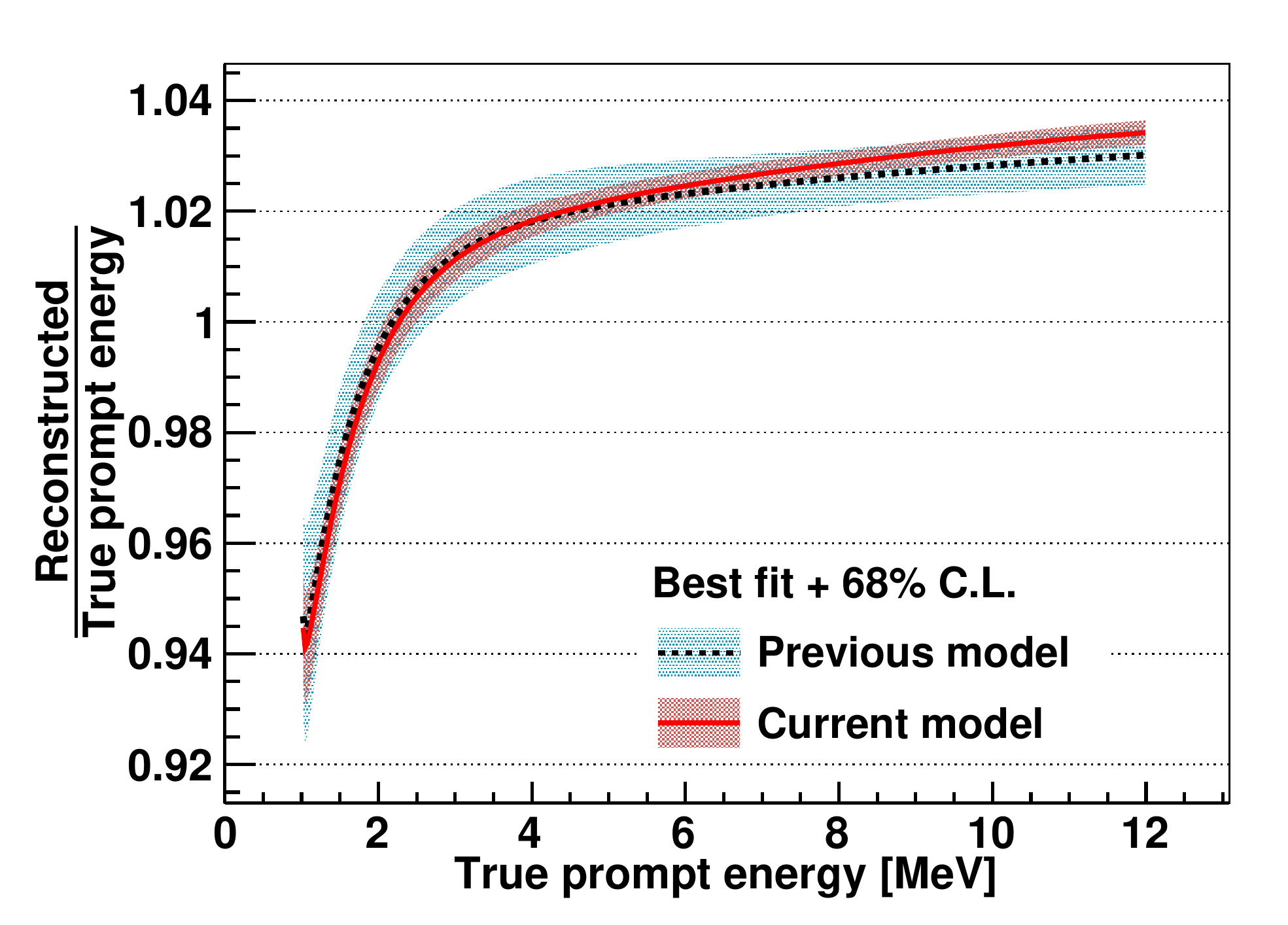}
\caption{\label{fig:positronNominal} Relationship between the reconstructed and true prompt energy,
which is the sum of positron kinetic energy and two 0.511~MeV annihilation $\gamma$~rays.
The previous model in Ref.\cite{DYBnGd2016} is shown for comparison~(colour on-line).}
\end{centering}
\end{figure}

The $\chi^2$~function in Eq.~\ref{eq:chi2} is minimized, resulting in a $\chi_{min}^2/$NDF = 70.6/(86-5) = 0.87.
The nonlinearity model of \positron, which is the prompt signal in the \nuebar~detection, is derived from the best-fit parameters, as shown in Fig.~\ref{fig:positronNominal}. The depicted uncertainty band corresponds to the models consistent with the calibration data within 68\% C.L.~(with a $\Delta \chi^2 < 5.89$ compared to $\chi_{min}^2$). A precision better than 0.5\% is achieved for prompt energies larger than 2~MeV.
The precision is limited by the systematic uncertainties associated with the $\gamma$~rays at energies below 3~MeV, and by the $^{12}$B statistics at higher energy.
As discussed in Sec.~\ref{positron}, the in-flight annihilation and the 3~$\gamma$~decay from o-Ps have a $<$0.1\% impact on the nonlinearity model. The best-fit full nonlinearity and scintillation nonlinearity of \electron, \positron and $\gamma$ are provided in the tabular form as Supplemental Materials~\cite{Supplemental}.

\begin{table*}[tb]
\begin{centering}

  \caption{Summary of free parameters in the $\chi^2$~minimization. The best-fit values and uncertainties are provided, and the correlation matrix is shown in Fig.~\ref{fig:correlationcoeff}.}
  \label{table:freeparameters}
  \small
  \begin{tabular}{c|c|c|c}
    \hline \hline
    Free parameter & Definition & Best-fit value                                     & Best-fit uncertainty \\ \hline
	$A$            & Absolute energy scale       & 1.013                                & 0.006 \\ \hline
	$kB$          & Birks' coefficient          & 15.2$\times 10^{-3}$~g/cm$^2$/MeV & 2.7$\times 10^{-3}$~g/cm$^2$/MeV \\ \hline
	$k_C$          & Absolute Cherenkov contribution   & 0.019    & 0.009  \\ \hline
	$\alpha$       & Amplitude of the electronics nonlinearity & 0.104                       & 0.002 \\ \hline
	$\tau$       & Decay constant of the electronics NL & 2.30~MeV          & 0.06~MeV \\
    \hline \hline
  \end{tabular}
 \end{centering}

\end{table*}

\begin{figure}[!h]
\begin{centering}
\includegraphics[width=.45\textwidth]{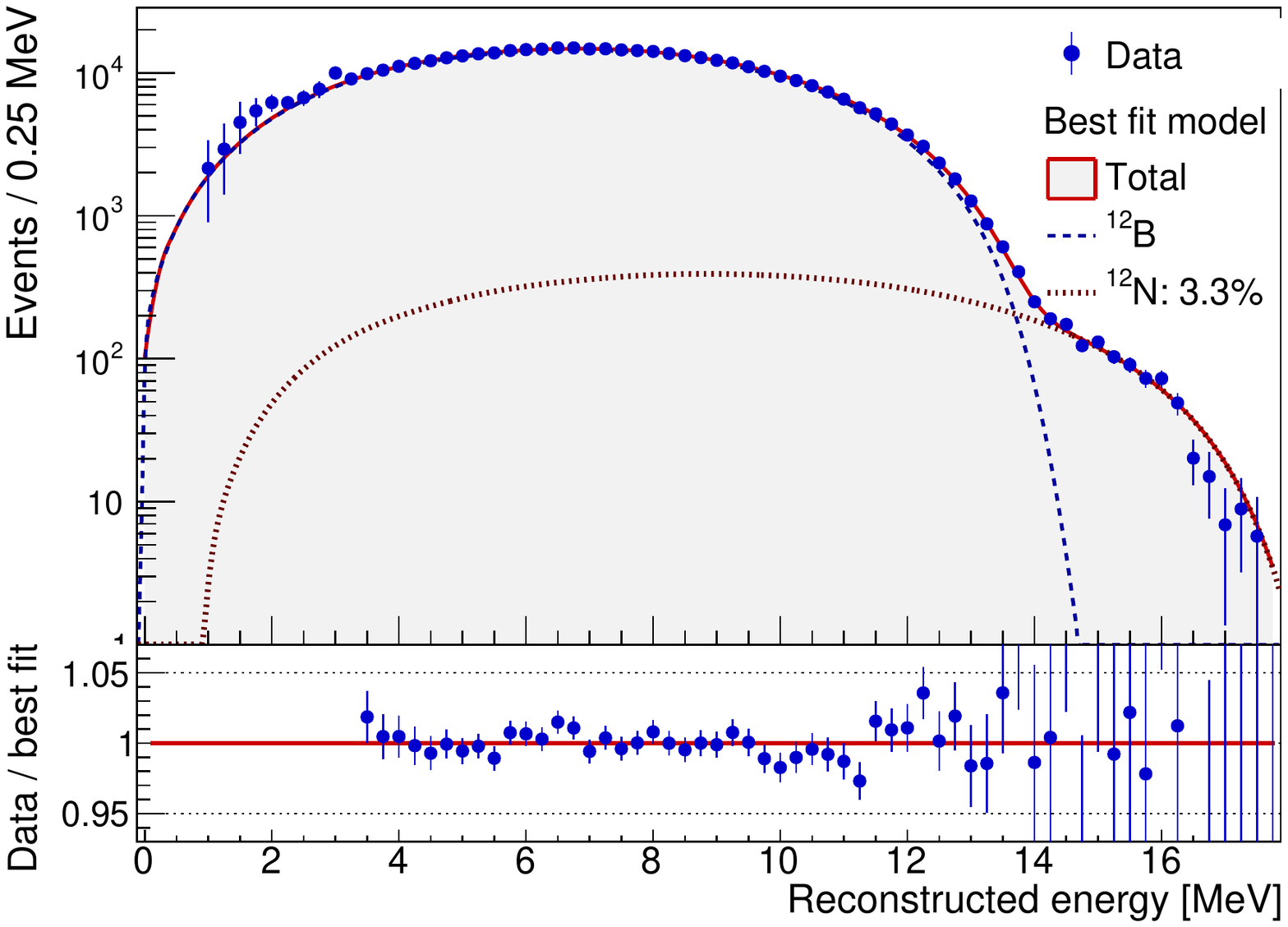}
\includegraphics[width=.45\textwidth]{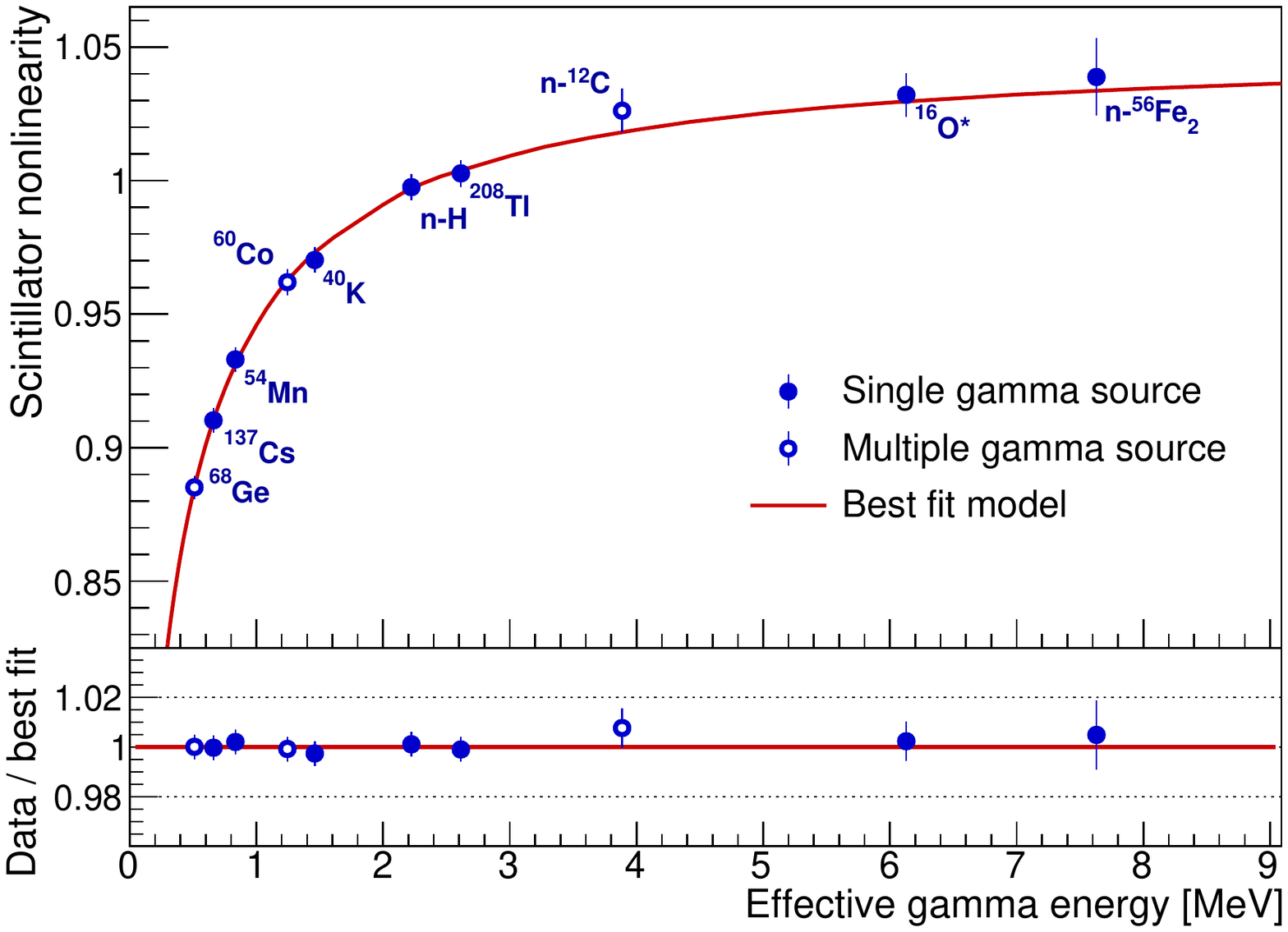}
\caption{\label{fig:B12Fit} Top: comparison of the reconstructed \B~energy spectra between the data
and the prediction with best-fit nuisance parameters.
Bottom: the estimated LS nonlinearity~(red line), and the measured from $\gamma$~rays. The error bars of the \B~spectrum represents the statistical uncertainty, while those of the $\gamma$~calibration points are taken from Table~\ref{table:gammaData}.}
\end{centering}
\end{figure}

\begin{figure}[!h]
\begin{centering}
\includegraphics[width=.45\textwidth]{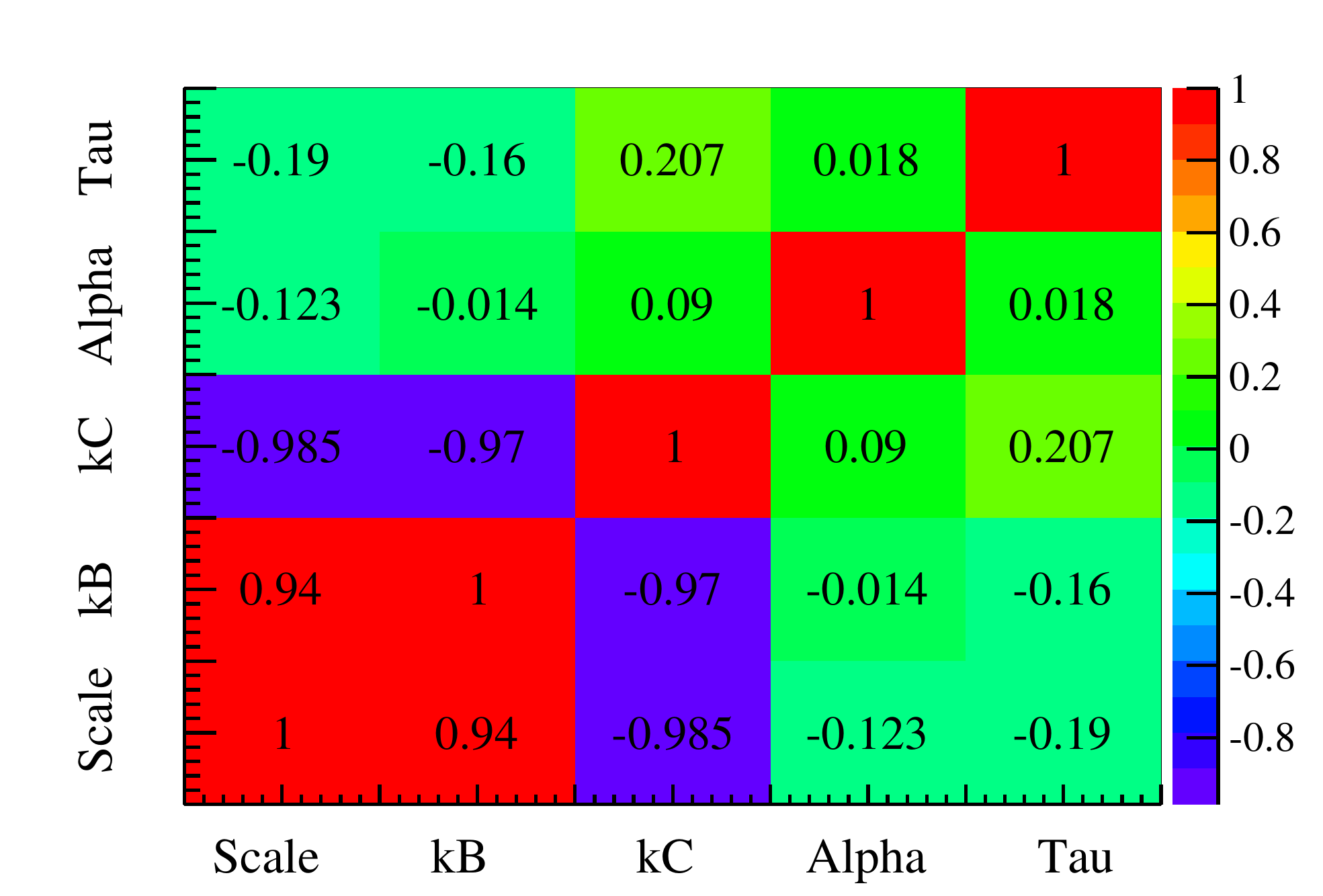}
\caption{\label{fig:correlationcoeff} Correlation coefficients of the five parameters.
The FADC measurement helps to decouple the LS and electronics nonlinearity. }
\end{centering}
\end{figure}

The energy nonlinearity model agrees with the calibration data well, as shown in Fig.~\ref{fig:B12Fit}. The values of best-fit parameter are provided in Table~\ref{table:freeparameters}, and
Fig.~\ref{fig:correlationcoeff} shows the correlation matrix of the five parameters. The Birks' coefficient, $kB$, has a large positive correlation with the absolute energy scale $A$~and a negative correlation with the Cherenkov contribution $k_C$. The correlation between the LS and electronics nonlinearity, $\alpha$~and $\tau$, is weak due to the constraints from the directly measured electronics nonlinearity. It is not practical to compare the best-fit values with those of other experiments due to the dependence on simulation parameters.

\begin{figure}[h]
\begin{centering}

\subfigure[]{\label{fig:positronNominalwoB12} \includegraphics[width=0.45\textwidth]{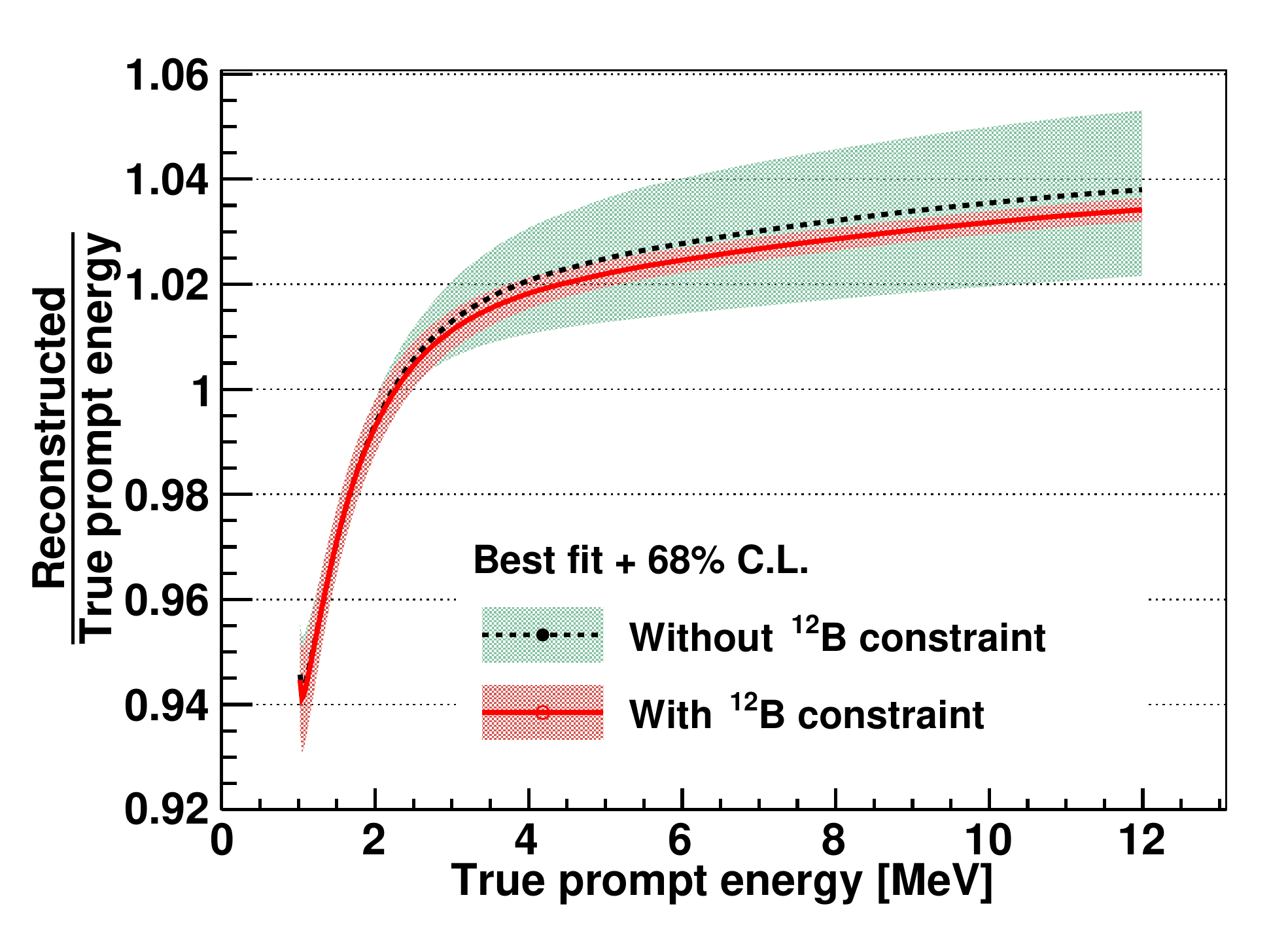}}
\subfigure[]{\label{fig:positronNominalwoFADC}\includegraphics[width=0.45\textwidth]{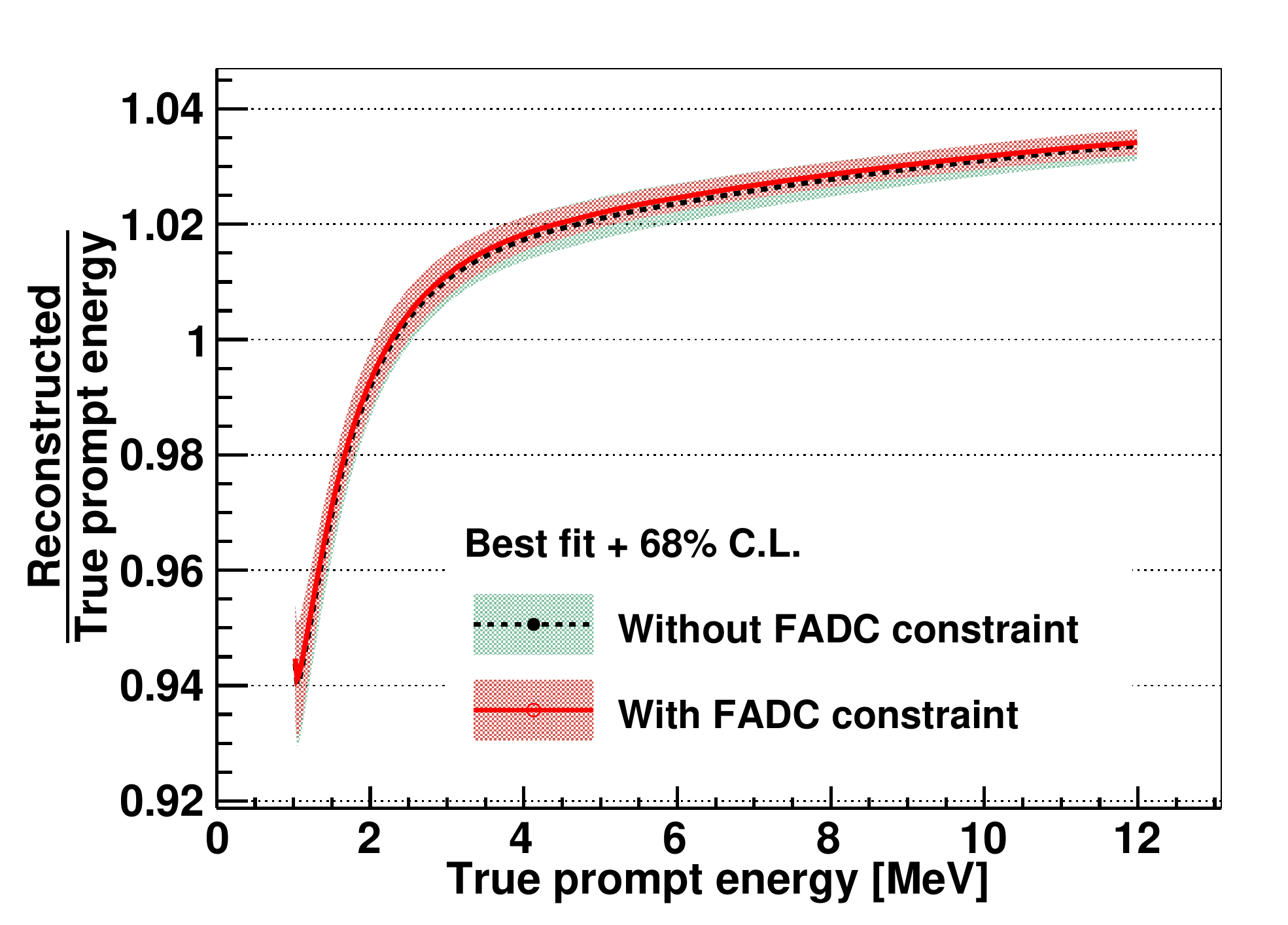}}
  
\caption{\label{fig:positronNominalwoB12FADC} Energy nonlinearity models of positrons without constraints from the \B~spectrum~(a) or the measured electronics nonlinearity~(b). }
\end{centering}
\end{figure}

Cross validations were performed by removing the $\gamma$~calibration points one by one, the measured electronics nonlinearity or the \B~from the fit.
Consistent results are found as shown in Fig.~\ref{fig:positronNominalwoB12FADC}.
Figure~\ref{fig:positronNominalwoB12} is the nonlinearity model without \B,
yielding three to eight times larger uncertainties for prompt energies greater than 4~MeV.
Although removing the measured electronics nonlinearity~(Fig.~\ref{fig:positronNominalwoFADC})~has little impact on the model precision, the correlation between the LS and electronics nonlinearity increases by a factor of 5 to 10.
The quenching curves produced by the different methods in Sec.~\ref{ElectronNL} are also tested in the fit, and the overall behavior of the \positron~nonlinearity is unchanged at all energies within 0.05\%.

\begin{figure}[!h]
\begin{centering}
\includegraphics[width=.45\textwidth]{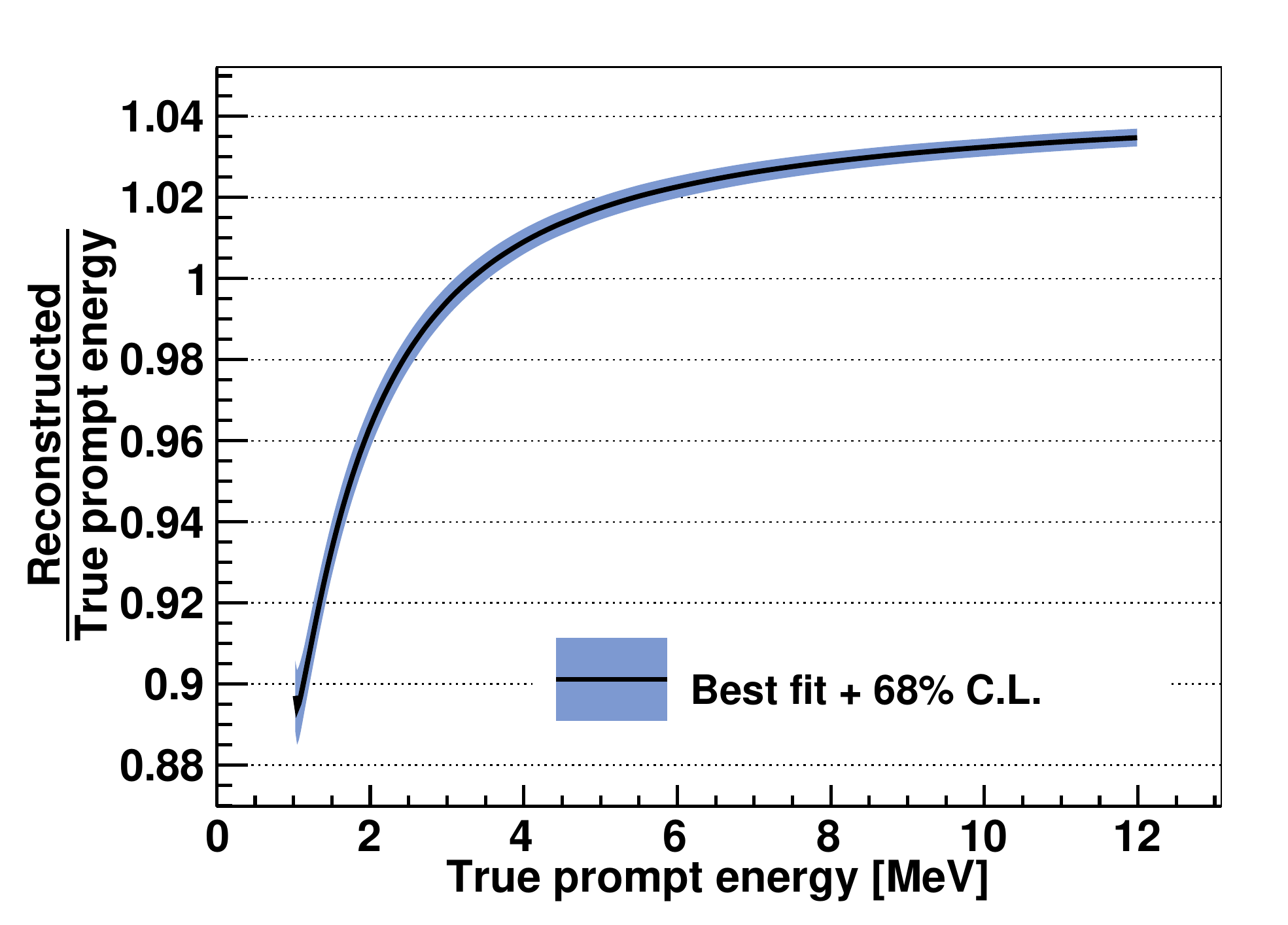}
\caption{\label{fig:positronNominalNew} Energy nonlinearity model of positrons with the electronics nonlinearity corrected at the channel level. The majority of the electronics nonlinearity is removed.}
\end{centering}
\end{figure}

The energy nonlinearity model of \positron~for the reconstructed energy with the electronics nonlinearity corrected at the single channel level is shown in Fig.~\ref{fig:positronNominalNew} and is provided as Supplemental Material~\cite{Supplemental}. The majority of electronics nonlinearity has been removed.
The scintillation nonlinearity derived from the two reconstructed energies converges to 0.1\%.

\section{Summary}
\label{Summary}

We present a precise energy calibration at Daya Bay. The nonlinearity in the existing electronics readout system is measured with a full FADC readout system. The measurements and systematic uncertainty estimates of the calibration data, including $\gamma$~rays and the \B~spectrum, are described in detail. Less than 0.5\% uncertainty in the energy nonlinearity calibration is achieved for positrons with kinetic energies greater than 1~MeV. There is no significant deviation in the nonlinearity between detectors, so the model is used for all eight ADs. A new reconstructed energy, with the electronics nonlinearity corrected at the channel level, has been implemented in the Daya Bay data processing.

\section{Acknowledgements}
\label{Acknowledgements}
Daya Bay is supported in part by the Ministry of Science and
Technology of China, the U.S. Department of Energy, the Chinese
Academy of Sciences, the CAS Center for Excellence in Particle
Physics, the National Natural Science Foundation of China, the
Guangdong provincial government, the Shenzhen municipal government,
the China General Nuclear Power Group, Key Laboratory of Particle and
Radiation Imaging (Tsinghua University), the Ministry of Education,
Key Laboratory of Particle Physics and Particle Irradiation (Shandong
University), the Ministry of Education, Shanghai Laboratory for
Particle Physics and Cosmology, the Research Grants Council of the
Hong Kong Special Administrative Region of China, the University
Development Fund of The University of Hong Kong, the MOE program for
Research of Excellence at National Taiwan University, National
Chiao-Tung University, and NSC fund support from Taiwan, the
U.S. National Science Foundation, the Alfred~P.~Sloan Foundation, the
Ministry of Education, Youth, and Sports of the Czech Republic,
the Charles University Research Centre UNCE,
the Joint Institute of Nuclear Research in Dubna, Russia, the National Commission of Scientific and
Technological Research of Chile, and the Tsinghua University
Initiative Scientific Research Program. We acknowledge Yellow River
Engineering Consulting Co., Ltd., and China Railway 15th Bureau Group
Co., Ltd., for building the underground laboratory. We are grateful
for the ongoing cooperation from the China General Nuclear Power Group
and China Light and Power Company.

\bibliographystyle{elsarticle-num}
\bibliography{EnergyCalibration}

\end{document}